\documentstyle [12pt]{article}
\thicklines
\newcommand{\nn}{\nonumber\\}

\newcommand{\be}{\begin{equation}}
\newcommand{\ee}{\end{equation}}
\newcommand{\bea}{\begin{eqnarray}}
\newcommand{\eea}{\end{eqnarray}}

\newcommand{\bfp}{{\bf p}}

\newcommand{\bfq}{{\bf q}}

\newcommand{\piN}{$\pi N$}

\newcommand{\NN}{$N\!N$}

\newcommand{\piNN}{$\pi N\!N$}

\newcommand{\bc}{\begin{center}}
\newcommand{\ec}{\end{center}}

\newcommand{\eqn}[1]{\label{#1}}
\newcommand{\eq}[1]{Eq.\ (\ref{#1})}
\newcommand{\eqs}[1]{Eqs.\ (\ref{#1})}
\newcommand{\fign}[1]{\label{#1}}
\newcommand{\fig}[1]{Fig.\ \ref{#1}}

\renewcommand{\theequation}{\arabic{equation}}

\begin{document}
\setcounter{section}{1}
\begin{picture}(100,20)(0,0)
\put (360,80){\hfill\begin{tabular}{l}
FIAS-R-224\\ September, 1993\end{tabular}
}\end{picture} \vspace{-1cm}

\large
\centerline{\bf Covariant three-body equations in{\boldmath $\phi^3$}
field theory\footnote{Accepted for publication in Nuclear Physics A}}
\normalsize

\bigskip
\bigskip

\centerline{A.\ N.\ Kvinikhidze}
\centerline{\em Mathematical Institute of Georgian Academy of Sciences}
\centerline{\em   Z.Rukhadze 1, 380093 Tbilisi, Georgia}

\vspace{0.6cm}

\centerline{B.\ Blankleider}
\centerline{\em School of Physical Sciences, Flinders University of South
Australia,}
\centerline{\em Bedford Park, S.A. 5042, Australia}

\vspace{.5cm}

\large
\centerline{\bf Abstract}
\normalsize
We derive four-dimensional relativistic three-body equations for the case
of a field theory with a three-point interaction vertex. These equations
describe the coupled $2\rightarrow 2$, $2\rightarrow 3$, and $3\rightarrow
3$ processes, and provide the means of calculating the kernel of the
$2\rightarrow 2$ Bethe-Salpeter equation. Our equations differ from all
previous formulations in two essential ways. Firstly, we have overcome the
overcounting problems inherent in earlier works. Secondly, we have retained
all possible two-body forces when one particle is a spectator. In this
respect, we show how it is necessary to also retain certain three-body
forces as these can give rise to (previously overlooked) two-body forces
when used in a $2\rightarrow 3$ process. The revealing of
such hidden two-body forces gives rise to a further novel feature of our
equations, namely, to the appearance of a number of subtraction terms. In
the case of the \piNN\ system, for example, the \NN\ potential involves a
subtraction term where two pions, exchanged between the nucleons, interact
with each other through the $\pi$-$\pi$ t-matrix. The necessity of an input
$\pi$-$\pi$ interaction is surprising and contrasts markedly with the
corresponding three-dimensional description of the \piNN\ system where no such
interaction explicitly appears. This illustrates the somewhat unexpected
result that the four-dimensional equations differ from the
three-dimensional ones even at the operator level.

\bigskip
\bigskip
\centerline{\bf I. INTRODUCTION}
\bigskip

Relativistic quantum field theory is the underlying basis of essentially all
models in both nuclear and particle physics. Yet our knowledge of how to
solve any but the most simple of models using such a field theory is
extremely limited. In this respect, it is useful to recall that, for the case
of one- and two-particle systems, the way to sum all possible perturbation
graphs is known, though only formally. In particular, the coupled problems of
single particle dressing, three-point vertex dressing, and $2\rightarrow 2$
scattering, is described in terms of the Dyson-Schwinger and Bethe-Salpeter
equations. It is important to recognize, however,  that despite the
"text-book" nature of these equations, they do not constitute a solution of
the one- and two-particle problem. The crucial point is that the essential
ingredient in these equations is the kernel of the $2\rightarrow 2$
Bethe-Salpeter equation, {\em but this kernel is not known}. Although one can
calculate contributions to this kernel up to some order in the coupling
constant, this is not a satisfactory method for strong interactions where it
is necessary to have a non-perturbative approach. In this paper we develop
integral equations for the coupled $2\rightarrow 2$, $2\rightarrow 3$, and
$3\rightarrow 3$ processes, which, in the case of negligible three-body forces,
also provide exact closed equations for the unknown Bethe-Salpeter kernel.

More generally, our goal is to formulate the field theoretic three-body
problem where one-, two-, and three-particle states are explicitly coupled,
and where three-body forces are taken as a given input (a reasonable first
approximation is to take them to be zero). Unfortunately, the formulation of
the four-dimensional three-body problem is not as straightforward as that for
the two-body case.  The basic problem is that "irreducibility", the essential
concept used to derive both the Dyson-Schwinger and Bethe-Salpeter equations,
cannot be applied to the case of three-particles {\em directly} - as we shall
see, quite a sophisticated additional analysis of diagram structure is
needed.  In fact, we find that in all the previous works on this subject,
such analyses of diagram structure are either incorrect, incomplete, or
missing altogether. Thus, despite the fundamental nature of the problem,
there does not appear to be any satisfactory formulation of the three-body
problem in four-dimensional relativistic field theory. The goal of this
paper, therefore, is to provide just such a formulation.

In order to develop equations for the field theoretic three-body system, we
also utilise the concept of irreducibility. A Feynman diagram is said to be
$r$-particle reducible if one can draw a continuous curve intersecting
exactly $r$ lines, at least one of which is an internal line, and each line
being crossed just once, so as to separate intial from final states. A
diagram that is not $r$-particle reducible, is said to be $r$-particle
irreducible. The Bethe-Salpeter equation then follows from the following
argument. Firstly, all $2\rightarrow 2$ Feynman diagrams are classified
according to their 2-particle irreducibility. The sum of all 2-particle
irreducible diagrams is defined to be the potential for the process. All
the rest of the diagrams have at least one 2-particle cut; moreover, it is
clear from the topology of such diagrams, that there must be a {\em unique}
left-most (or right-most) 2-particle cut. To the left of this left-most cut
is a 2-particle irreducible potential, and the Bethe-Salpeter equation
follows immediately.

In a pioneering work, Taylor \cite{Taylor} applied such a procedure to the
three-body system. He recognised that, unlike the case for 2-particle cuts,
3-particle cuts may not be unique. Particular attention was paid to the
case illustrated in \fig{two_cut}(a). Here is shown a diagram that is 2-
and 3-particle reducible, and where it is possible to cut three internal
lines in two different ways, neither of which precedes the other. It was
recognised that to eliminate such non-unique {\em internal} 3-particle cuts,
one should remove the two-particle cuts first. Formalizing these considerations
into his "last cut" lemma, Taylor used the scheme where $3$-particle cuts are
exposed only in $2$-irreducible amplitudes, to derive four-dimensional
relativistic equations for the field theoretic three-body problem.
\thicklines

\begin{figure}[t]
\vspace{2cm}
\caption{\fign{two_cut}
Diagrams where there is no unique right-most
3-particle cut. (a) 2-particle reducible
example, (b) 2-particle irreducible example.}
\begin{picture}(0,10)(-60,-10)
\put (-10,0){
\begin{picture}(100,100)(0,0)
\put (-50,90){(a)}
\put (5,105){$_{2'}$}
\put (5,90){$_{3'}$}
\put (5,75){$_{1'}$}
\put (130,105){$_{2}$}
\put (130,75){$_{1}$}
\put (20,105.){\line(1,0){100.}}
\put (20,75){\line(1,0){100.}}
\put (40,90){\oval(15,30)[r]}
\put (100,90){\oval(15,30)[l]}
\put (20,90){\line(1,0){20}}
\multiput(40,75)(30,0){3}{\line(0,1){30.}}
\multiput(70,90)(-2,2){16}{\circle*{.05}}
\multiput(70,90)(2,2){16}{\circle*{.05}}
\multiput(70,90)(-2,-2){16}{\circle*{.05}}
\multiput(70,90)(2,-2){16}{\circle*{.05}}
\end{picture}}

\put (210,0){
\begin{picture}(100,100)(0,0)
\put (-50,90){(b)}
\put (5,105){$_{2'}$}
\put (5,90){$_{3'}$}
\put (5,75){$_{1'}$}
\put (110,105){$_{2}$}
\put (110,75){$_{1}$}
\put (20,105.){\line(1,0){80.}}
\put (20,75){\line(1,0){80.}}
\put (20,90){\line(4,1){60.}}
\put (80,75){\line(-1,1){30.}}
\multiput(105,60)(-2,2){32}{\circle*{.05}}
\multiput(40,60)(2,2){32}{\circle*{.05}}
\end{picture}}

\end{picture}
\end{figure}
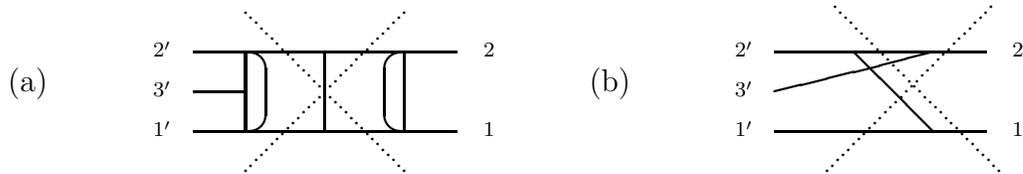

However, no special attention was given  to the case illustrated in
\fig{two_cut}(b). Shown is a diagram for the $2\rightarrow 3$ process that is
3-particle reducible, but 2-particle irreducible. Shown also are two 3-particle
cuts, neither of which is to the right of the other, i.e.\ there is no unique
right-most 3-particle cut, despite the fact that two-particle cuts have been
removed. This lack of uniqueness in the cutting procedure, necessarily
involving at least one {\em external} particle, does not seem to have been
taken into account by Taylor, or in a number of subsequent works that were
based on his "last cut" lemma [2-5]; as a result, the three-body equations
developed in all the above mentioned works suffer from the problem of
overcounting.

The only previous work where this type of non-uniqueness problem was
addressed, is the one by Tucciarone \cite{Tucciarone}. As a result, he
developed a description of the four-dimensional three-body problem where
there is no overcounting of diagrams. However, his equations were not
expressed in terms of subsystem two-body amplitudes, rather, an explicit
perturbation expansion for the inhomogeneous terms is needed. This makes his
equations unsatisfactory from a practical point of view.

It should be noted, that the non-uniqueness difficulty does not arise for
$3\rightarrow 3$ processes where coupling to two-particle states does not
take place, as for example in formulations of the relativistic
four-dimensional three-nucleon problem \cite{Stojanov}. This problem also
does not arise in time-ordered perturbation theory since, for example, the
two cuts in \fig{two_cut}(b) would correspond to the two possible time
orderings of the right-most vertices, both of which need to be taken into
account.

In the present work, we restrict ourselves to the case of a $\phi^3$ field
theory. This is the simplest case having wide applicability; however, the
strategy we shall adopt is also applicable to other forms of the interaction.
In particular, we overcome the non-uniqueness problem by a special procedure
where, in cases like \fig{two_cut}(b), one of the two rightmost vertices is
"pulled out" further to the right. Our equations therefore do not suffer from
the overcounting problem inherent in the above mentioned works.

A second, but equally important goal of this paper, is to take into account
as many Feynman diagrams as possible, consistent with pair-like interactions.
That this is not an altogether obvious task, is illustrated in \fig{3b}.
\fig{3b}(a) shows a connected three-body force for the $3\rightarrow 3$
process. Yet if we join any two initial or final legs into a single leg, i.e.
if there is absorption through a $\phi^3$ vertex, then the resulting diagram
can be represented through a pair-like interaction, as shown in \fig{3b}(b).
Since the equations we seek couple $3\rightarrow 3$ to $3\leftrightarrow 2$
processes, \fig{3b} illustrates how careful one must be in neglecting
three-body forces, otherwise one can inadvertently neglect important two-body
contributions as well. This subtlety does not appear to have been noticed
before, as all the References [1-5] have neglected three-body forces without
comment. Note, that in an otherwise correct formulation, the neglect of such
three-body forces, leading to the undercounting of pair-like interactions,
would result in equations that have the wrong three-dimensional limit; thus,
they would have neither the correct non-relativistic limit, nor would they
satisfy three-body unitarity.
\thicklines
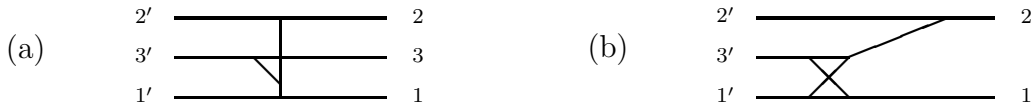
\begin{figure}[t]
\vspace{1.5cm}
\caption{\fign{3b}
(a) A three-body force in the $3\rightarrow 3$ process.
(b) The same three-body force, after particle $1$ has absorbed particle $3$,
reveals a two-body process.}
\begin{picture}(0,10)(20,-10)
\put (0,0){ \begin{picture}(100,100)
\put (20,80){(a)}\end{picture}
\begin{picture}(100,100)(40,10)
\put (5,105){$_{2'}$}
\put (5,90){$_{3'}$}
\put (5,75){$_{1'}$}
\put (110,105){$_{2}$}
\put (110,90){$_{3}$}
\put (110,75){$_{1}$}
\put (20,105.){\line(1,0){80.}}
\put (20,90){\line(1,0){80.}}
\put (50,90){\line(1,-1){10}}
\put (20,75){\line(1,0){80.}}
\put (60,75){\line(0,1){30.}}
\end{picture}}

\put (220,0){ \begin{picture}(100,100)
\put (20,80){(b)}\end{picture}
\begin{picture}(100,100)(40,10)
\put (5,105){$_{2'}$}
\put (5,90){$_{3'}$}
\put (5,75){$_{1'}$}
\put (120,105){$_{2}$}
\put (120,75){$_{1}$}
\put (20,105.){\line(1,0){90.}}
\put (20,90){\line(1,0){35.}}
\put (20,75){\line(1,0){90.}}
\put (55,90){\line(5,2){38.}}
\put (40,75){\line(1,1){15.}}
\put (40,90){\line(1,-1){15.}}
\end{picture}}

\end{picture}
\end{figure}

In this work we present three-body equations that suffer neither from the
overcounting nor the undercounting problems discussed above.  A detailed
derivation is presented for the case of three identical Bosons.  The case
of identical particles was chosen because it has all the complications of
the most general case. It is also the most difficult case that can arise in
practice, where, in particular, pair-like interactions are less
straightforward to reveal.  

We also present the equations for the case of the \piNN\ system where only two
of the particles are identical. This system has commanded much attention in the
literature although mainly in the three-dimensional sector \cite{Garcilazo}; in
this respect, we note that the recently derived convolution equations of Ref.\
\cite{KB2} can be viewed as the three-dimensional limit of the four-dimensional
equations we present in this paper.

A novel feature of our equations is the presence of explicit subtraction
terms necessary to avoid overcounting. In the \piNN\ equations, for
example, one of the subtraction terms is the two-pion exchange \NN\
potential where the exchanged pions interact with each other. As the
corresponding three-dimensional equations have only the \piN\ and \NN\
interactions as input, it is remarkable that a consistent four-dimensional
description of the \piNN\ system also demands, as input, the  $\pi$-$\pi$
interaction.

\bigskip
\bigskip \newpage

\setcounter{section}{2}
\setcounter{equation}{0}
\centerline{\bf II. DERIVATION}
\bigskip
For the $m\rightarrow n$ process the momentum space Green's function is
defined by
\bea
\lefteqn{
{\cal G}(p'_1 \dots p'_n,p_1 \dots p_m) = \int \prod_{j=1}^n d^4x'_j \,
\prod_{i=1}^m \, d^4x_i  \, \mbox{exp}\left[ i\sum_{j=1}^n x'_jp'_j -
i\sum_{i=1}^m x_ip_i\right] } \hspace{3cm} \nn
&&\frac{<0|T[\phi_{1'}(x'_1)\dots\phi_{n'}(x'_n)
\phi_1(x_1)\dots\phi_m(x_m)S]|0>} {<0|S|0>}. \hspace{1cm}   \eqn{GG}
\eea
where the S-matrix $S$ is given in terms of the interaction Lagrangian
${\cal L}_I$ by
\be
S=T\mbox{exp}\left[i\int d^4x\,{\cal L}_I(x) \right].
\ee

For clarity of presentation, we shall usually replace momenta by their labels
in
an obvious way; thus we write ${\cal G}(1' \dots n',1 \dots m)$ instead of
${\cal
G}(p'_1 \dots p'_n,p_1 \dots p_m)$. For momentum conservation $\delta$
functions, we use a convention where, for example, $\delta(1'2'3',12)$ denotes
the function $\delta(p'_1 + p'_2 + p'_3 - p_1 - p_2)$. For the case
$n=m$, it will also be useful to define ${\cal G}_0(1' \dots n',1 \dots n)$ as
that part of ${\cal G}(1'\dots n',1 \dots n)$ corresponding graphically to $n$
completely disconnected lines connecting initial and final states.

For specific one-, two- and three-particle Green's functions we shall write
respectively  \bea
&&g(1',1)  = {\cal G}(1',1)                  \eqn{g}   \\
&&D(1'2',12)  = {\cal G}(1'2',12)            \eqn{D}   \\
&&G(1'2'3',123)  = {\cal G}(1'2'3',123)   .  \eqn{G}
\eea
We shall also need corresponding disconnected propagators where the
momentum conserving $\delta$ functions have been factored out; in these cases
we
write
\bea
&&g(1',1)  = d(1) \delta(1',1)                                  \eqn{d}  \\
&&D_0(1'2',12) = d_1(1) d_2(2) \delta(1',1) \delta(2',2)        \eqn{dd} \\
&&G_0(1'2'3',123) = d_1(1) d_2(2) d_3(3)
                    \delta(1',1) \delta(2',2) \delta(3',3)  .   \eqn{ddd}
\eea
For non-identical particles, we have that
\bea
&&D_0(1'2',12)  = {\cal G}_0(1'2',12)         \eqn{D_0} \\
&&G_0(1'2'3',123)  = {\cal G}_0(1'2'3',123) . \eqn{G_0}
\eea
 We shall work also with t-matrix amplitudes defined in terms of the above
Green's functions by the relation
.\be
{\cal T}(1' \dots n',1 \dots m) = \prod_{j=1}^n d_{j'}^{-1}(j')
{\cal G}'(1' \dots n',1 \dots m)   \prod_{i=1}^m d_{i}^{-1}(i)     \eqn{TT}
\ee
where ${\cal G}'={\cal G}$ if $m\ne n$, and ${\cal G}'={\cal G}-{\cal G}_0$ if
$m=n$. Just as for propagators, we shall use specific symbols for the
amplitudes
of particular processes. Namely, we use symbol $f$ for the dressed
$1\rightarrow
2$ vertex,   $t$ for the $2\rightarrow 2$ process, $\Gamma$ for the
$1\rightarrow 3$ process, $M$ for the $2\rightarrow 3$ process, and $T$ for the
$3\rightarrow 3$ process. Thus \bea
&&f(1'2',1) = {\cal T}(1'2',1) \eqn{f_def}      \\
&&t(1'2',12) = {\cal T}(1'2',12) \eqn{t_def}          \\
&&\Gamma(1'2'3',1) = {\cal T}(1'2'3',1)  \eqn{Gamma_def}  \\
&&M(1'2'3',12) = {\cal T}(1'2'3',12) \eqn{M_def}      \\
&&T(1'2'3',123) = {\cal T}(1'2'3',123) \eqn{T_def} .
\eea
The basic three-point vertex, without a momentum conserving $\delta$ function
is
defined as $h$, thus
\be
f(1'2',1) = h(1'2',1) \delta(1'2',1).
\ee
For the disconnected $2\rightarrow 3$ process it will be sufficient to define
\bea
F_1(1'2'3',12) &=& f(1'3',1) d_2^{-1}(2) \delta(2',2)    \eqn{F_i}   \\
F_2(1'2'3',12) &=& f(2'3',2) d_1^{-1}(1) \delta(1',1)  .
\eea

To save on notation further, we shall usually suppress particle labels
completely. Thus we denote any quantity $A(1'\ldots n',1\ldots m)$ simply
by $A$. Note that the order of labels in this definition is important.
As occasionally a different order of labels will be needed, it will be useful
to introduce the left and right "particle exchange" operators $L_{ij}$ and
$R_{ij}$; when acting on any quantity $A$, these operators exchange the
$i$th and $j$th labels on the left and right set of external legs,
respectively.
For example, we have that
\bea
F_1 &=& F_1(1'2'3',12), \\
R_{12} F_1 &=& F_1(1'2'3',21), \\
L_{13} F_1 &=& F_1(3'2'1',12).
\eea

For reversed processes we use a {\em bar} notation; thus, $\bar{f}$ is the
three-particle vertex for absorption, and $\bar{M}$ is the amplitude for the
$3\rightarrow 2$ process.

We shall also write integrals in symbolic form. For example, for any two
quantities $B$ and $A$, describing processes $m\rightarrow k$ and
$k\rightarrow n$ respectively, we define $AB$ by
\be
AB(p'_1\ldots p'_n,p_1\ldots p_m) \equiv
\int d^4p''_1\ldots d^4p''_k \,A(p'_1\ldots p'_n,p''_1\ldots p''_k)
B(p''_1\ldots p''_k,p_1\ldots p_m) . \ee
Thus an equation of the form
\be
C(p'_1\ldots p'_n,p_1\ldots p_m) = \int d^4p''_1\ldots d^4p''_k \,A(p'_1\ldots
p'_n,p''_1\ldots p''_k) B(p''_1\ldots p''_k,p_1\ldots p_m)
\ee
will be written symbolically as just $C=AB$.

\bigskip
\bigskip 
\centerline{\bf A. Identical particle system}
\bigskip

We first consider the case of identical particles since this contains all the
complications of the most general case. To be definite, we assume that the
identical particles are Bosons and that the underlying interaction Lagrangian
is
given by
\be
{\cal L}_I(x) = \frac{\lambda}{3!}  \phi^3(x).      \eqn{L_I}
\ee
Here $\lambda$ is the coupling constant and the factor $1/3!$ is chosen to
cancel factors arising from the permutation symmetry of the fields in
contractions when using Wick's theorem. Having defined the model in which we
shall work, it will now be assumed that all three-point vertices and particle
propagators are dressed, so we  consider the  skeleton Feynman diagrams of a
given process.

For identical Bosons, the Green's function ${\cal G}$ of \eq{GG} is symmetric
under the interchange of either initial or final labels.  In this case we still
retain all the above definitions, so the full propagators $D$, $G$, and
t-matrices $f$, $\Gamma$, $t$, $M$, and $T$ are likewise symmetric in their
initial and final labels. Note, however, that the free propagators $D_0$ and
$G_0$ do not have a specific symmetry, and the function $F_i$ is symmetric
under
the interchange of $i$ and $3$ momentum labels, but does not have a specific
symmetry under the exchange of labels $1$ and $2$.

Quantities without a specific symmetry, will sometimes need to be symmetrized
explicitly. Thus, for example, we have that
\bea
&&\sum_P D_0(1'2',12)  = {\cal G}_0(1'2',12)        \eqn{sum_D_0} \\
&&\sum_P G_0(1'2'3',123)  = {\cal G}_0(1'2'3',123)  \eqn{sum_G_0}
\eea
where the sum is over all permutations of either the initial or final momenta.
More generally, we shall indicate sums over permutations of initial momenta by
the letter $R$ (right), and sums over permutations of final momenta by the
letter $L$ (left); however, we use the letter $P$ whenever it makes no
difference which sum, $R$ or $L$, is taken. Quantities symmetrized in one of
these ways are indicated by the appropriate superscript. Thus, for example, if
$A$ has three initial and final legs, then
\bea
A^R(1'2'3',123) = A(1'2'3',123) + A(1'2'3',213) + A(1'2'3',321) + \dots
\eqn{A_R} \\
A^L(1'2'3',123) = A(1'2'3',123) + A(2'1'3',123) + A(3'2'1',123) + \dots
\eqn{A_L}
\eea \newpage\noindent
with similar expressions holding for $A$ having any number of legs.
In general, we can write
\bea
A^R &\equiv& \sum_R A  \nn
A^L &\equiv& \sum_L A \nn
A^P &\equiv& \sum_P A = A^R = A^L . \nonumber
\eea
In the case of three-particle states, the symbols $R_c$, $L_c$, and $P_c$ will
be used to indicate sums over the corresponding {\em cyclic} permutations of
particles.

We shall utilise the idea of irreducibility described in the Introduction. The
maximal irreducibility of an amplitude will be indicated by a superscript in
round brackets; thus by $T^{(r)}$ we mean the sum of all possible $3\rightarrow
3$ skeleton Feynman graphs that are simultaneously $1$-, $2$-, \dots,
$r$-particle irreducible.

The symmetry property of identical particle amplitudes necessitates the
inclusion of various counting factors, as well as sums over permutations of
particle labels whenever amplitudes do not already have the required symmetry.
In the Appendix, we discuss the origin of such counting factors, and give
particular examples of how to symmetrize amplitudes. As a result, we shall
utilise such counting factors and symmetry operations in the discussion below,
but with only a minimal comment.

\bigskip
\bigskip 
\centerline{\em 1. The $3\rightarrow 3$ process}
\bigskip

We begin by examining the structure of the full amplitude $T$ for the
$3\rightarrow 3$ process. We can expose the one-particle cut through
the equation
\be
T = T^{(1)} + \Gamma^{(1)} g \bar{\Gamma}^{(1)}    \eqn{T}
\ee
which is illustrated diagrammatically in \fig{figT}.
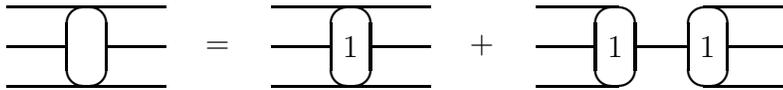
\begin{figure}[b]
\vspace{2cm}
\caption{
\fign{figT}
Graphical representation of the equation
$T=T^{(1)}+\Gamma^{(1)}g\bar{\Gamma}^{(1)}$.}

\thicklines
\begin{picture}(10,0)(-20,20)

\put (0,0){\begin{picture}(100,100)
\multiput (20,75.)(0,30){2}{\line(1,0){60.}}
\put (50,90){\oval(15,30)}
\multiput (20,90)(37.5,0){2}{\line(1,0){22.5}}
\end{picture}}

\put (100,0){\begin{picture}(100,100)
\put (-5,87.5){=}
\multiput (20,75.)(0,30){2}{\line(1,0){60.}}
\put (50,90){\oval(15,30)}
\put (47,86){1}
\multiput (20,90)(37.5,0){2}{\line(1,0){22.5}}
\end{picture}}

\put (200,0){\begin{picture}(100,100)
\put (-5,87.5){+}
\multiput (50,90)(35,0){2}{\oval(15,30)}
\multiput (47,86)(35,0){2}{1}
\multiput (20,90)(72.5,0){2}{\line(1,0){22.5}}
\multiput (20,75)(0,30){2}{\line(1,0){30.}}
\multiput (85,75)(0,30){2}{\line(1,0){30.}}
\put (57.5,90){\line(1,0){20}}
\end{picture}}

\end{picture}
\end{figure}
In the same way, we can expose the two-particle cut in $T^{(1)}$
\bea
T^{(1)} &=& T^{(2)} + M^{(2)} [\frac{1}{2} D_0 + \frac{1}{2} D_0
t^{(1)} \frac{1}{2}  D_0 ] \bar{M}^{(2)} \nn
 &=& T^{(2)} + \frac{1}{4} M^{(2)} [D_0^P + D_0 t^{(1)} D_0 ] \bar{M}^{(2)}.
                                                          \eqn{T_1}
\eea
Note that the factors of $1/2$ in the first of the above two expressions are
associated with the symmetric nature of the amplitudes $M^{(2)}$ and $t^{(1)}$
under the interchange of the exposed two particles; more generally, exposing
$n$ particles between appropriately symmetric amplitudes will result in a
counting factor of $1/n!$ - see the Appendix for details. By exposing the first
vertex in amplitude $\Gamma^{(1)}$, we can write it as
\be
\Gamma^{(1)} = \frac{1}{2} M^{(2)} D_0 f              \eqn{Gamma_1}
\ee
\begin{figure}[b]
\vspace{3cm}
\caption{
\fign{Gam} Graphical representation of the equation
$\Gamma^{(1)} = \frac{1}{2} M^{(2)} D_0 f$}
\thicklines
\begin{picture}(10,50)(-90,-20)

\put (0,0){\begin{picture}(100,100)
\multiput (20,75.)(0,30){2}{\line(1,0){27.5}}
\put (50,90){\oval(15,30)}
\put (47,86){1}
\multiput (20,90)(37.5,0){2}{\line(1,0){22.5}}
\end{picture}}

\put (100,0){\begin{picture}(100,100)
\put (-5,87.5){=}
\multiput (20,75.)(0,30){2}{\line(1,0){27.5}}
\put (50,90){\oval(15,30)}
\put (47,86){2}
\multiput (20,90)(62.5,0){2}{\line(1,0){22.5}}
\put (55,104){\line(2,-1){28}}
\put (55,76){\line(2,1){28}}
\end{picture}}

\end{picture}
\end{figure}
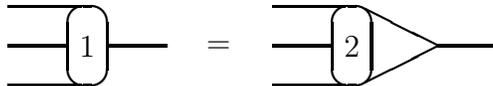
\noindent
which is illustrated in \fig{Gam}. Note that because all our vertices are
dressed from the beginning, the $2\rightarrow 3$ amplitude in \eq{Gamma_1} is 1
{\em and} 2-particle irreducible. Substituting \eq{T_1} into \eq{T} and using
\eq{Gamma_1} we obtain
\be
T = T^{(2)} +  \frac{1}{4} M^{(2)} D \bar{M}^{(2)}.      \eqn{T_a}
\ee
where we have used the fact that
\be
t =  t^{(1)} + f g \bar{f}                                \eqn{t}
\ee
and
\be
D = D_0^P +  D_0 t D_0.                          \eqn{DD}
\ee
In \eq{DD} we can treat $t$ in two ways. In order to compare with some
earlier works we would treat $t$ as a given input to the equations. In the
second way we would expose three-particle states in $t$ and thereby obtain
coupled equations for the $3\rightarrow 3$ and $2 \rightarrow 2$ processes.
Let us assume for the moment that $t$ is known. Then the only unknowns
in \eq{T_a} are the amplitudes $T^{(2)}$ and $M^{(2)}$.

For the amplitude $T^{(2)}$ any three-particle cut is unique as there are
no two-particles states in this amplitude. This means that we can immediately
write the three-particle Bethe-Salpeter equation for this amplitude
\be
T^{(2)} = K + \frac{1}{3!}K G_0 T^{(2)}    \eqn{T_12}
\ee
where $K$ is the sum of all $3\rightarrow 3$ diagrams that are simultaneously
1-, 2- and 3-particle irreducible. We write the disconnected part of $K$,
indicated by subscript $d$, in terms of 1- and 2-particle irreducible two-body
potentials $v$ $(\equiv t^{(2)})$ \be
K_d(1'2'3',123) = \sum_{L_cR_c} v(2'3',23) d^{-1}(1) \delta(1',1)   \eqn{K_d}
\ee
where $L_c$ and $R_c$ indicate that the sum is taken over cyclic permutations
of the left labels  $(1'2'3')$ and right labels $(123)$ respectively (note that
the sum is restricted to cyclic permutations because the potentials $v$ are
already symmetric in their labels). Note that
\be
t^{(1)} = v + \frac{1}{2} v D_0 t^{(1)}.     \eqn{t_(1)}
\ee
Defining $V_i$ by
\bea
V_1(1'2'3',123) &=& v(1'3',13) d^{-1}(2) \delta(2',2) \\
V_2(1'2'3',123) &=& v(2'3',23) d^{-1}(1) \delta(1',1) \\
V_3(1'2'3',123) &=& v(1'2',12) d^{-1}(3) \delta(3',3) ,  \eqn{v_3}
\eea
we have that
\be
K_d = \sum_{P_c} (V_1 + V_2 + V_3)   \eqn{K_d_short}
\ee
where it makes no difference over which labels, left or right, the cyclic
permutations are taken. The subscript $i$ of $V_i$ can be considered as
labelling the interacting pair. In this respect our convention throughout this
paper shall be as in the above equations, i.e., $1$ labels the pair $(13)$,
$2$ labels the pair $(23)$, and $3$ labels the pair $(12)$. Note that the
definitions of $F_i$ given earlier in \eqs{F_i} are consistent
with this convention. Further, we have that \be K = \sum_{P_c} (V_1+V_2+V_3) +
K_c
\ee
where $K_c$ is the connected part of $K$, and as such, is a three-body force.
Defining
\be
V = \frac{1}{2}(V_1+V_2+V_3) + \frac{1}{6}K_c  \eqn{V}
\ee
so that
\be
K = \sum_P V,
\ee
\eq{T_12} can be written as
\be
T^{(2)} = \sum_P V + V G_0 T^{(2)}  .  \eqn{T_12_P}
\ee
In obtaining \eq{T_12_P} we made use of the fact that $T^{(2)}$ is symmetric;
it is therefore gratifying to see that this derived equation now implies the
symmetry of $T^{(2)}$ explicitly.

In this work, we would like to be able to neglect three-body forces, but as
discussed in the Introduction, this needs to be done very carefully. In
\eq{T_12_P},  we can in fact neglect $K_c$ in the expression for $V$; however,
as we shall see shortly, this can be done only because there are no two-body
states in the amplitude $T^{(2)}$. As it stands, \eq{T_12_P} does not have a
compact kernel so a Faddeev rearrangement needs to be performed. To facilitate
this, one can recast \eq{T_12_P} into a more familiar form (for the scattering
of three non-identical particles) by defining the t-matrix $\tilde{T}$ through
the equation \be \tilde{T} = V + V G_0 \tilde{T}      \eqn{T_tilde}
\ee
where
\be
T^{(2)} = \sum_P \tilde{T} .
\ee
\eq{T_tilde} can then be transformed to give equations of standard Faddeev
form.
Although the introduction of $\tilde{T}$ is the familiar way of handling
\eq{T_12_P}, we show that in fact one can work with \eq{T_12_P} directly. To do
this, it is more straightforward to work with Green's functions. We thus write
\bea
G^{(2)} &=&  G_0^P + G_0  T^{(2)} G_0,        \eqn{G_12_T} \\
         &=&  G_0^P + G_0  V G^{(2)} .        \eqn{G_12}
\eea
Both these equations for $G^{(2)}$ and \eq{T_12_P} for $T^{(2)}$ differ from
the corresponding ones for non-identical particles only in the symmetrization
of
the inhomogeneous term; moreover, in the equations for  $G^{(2)}$, this
symmetrization is of the free Green's function $G_0$ and is thus particularly
simple. Without the $K_c$ term, \eq{G_12} is given by
\be
G^{(2)} =  G_0^P + \frac{1}{2} G_0 (V_1+V_2+V_3) G^{(2)} .  \eqn{G_12_1}
\ee
We define the components $T_i$ by
\be
T_i G_0 = \frac{1}{2} V_i G^{(2)}        \eqn{T_i_def}
\ee
so that
\be
G^{(2)} =  G_0^P + \sum_{i=1}^{3} G_0 T_i G_0. \eqn{G_12_2}
\ee
and
\be
T^{(2)} = T_1 + T_2 + T_3 .        \eqn{T_2_sum}
\ee
Using \eq{G_12_2} in \eq{T_i_def}, we obtain
\be
T_i G_0 = \frac{1}{2} V_i G_0^P + \frac{1}{2} V_i \sum_{j=1}^{3} G_0
T_j G_0,
\ee
\be
(1 - \frac{1}{2}V_iG_0) T_i = \frac{1}{2} V_i G_0^P G_0^{-1}
+ \frac{1}{2}V_i G_0 \sum_{j\ne i} T_j ,
\ee
and thus the Faddeev equations for the components $T_i$,
\be
T_i = \frac{1}{2} t_i^R + \frac{1}{2} t_i G_0 \sum_{j\ne i} T_j  \eqn{T_i}
\ee
where
\be
t_i = V_i + \frac{1}{2} V_i G_0 t_i .                   \eqn{t_i}
\ee
Note that the inhomogeneous term of \eq{T_i} involves a sum over right
(but not left!) permutations.
Comparing with \eq{t_(1)}, we see that the $t_i$ are explicitly given by
\bea
t_1(1'2'3',123) &=& t^{(1)}(1'3',13) d^{-1}(2) \delta(2',2) \\
t_2(1'2'3',123) &=& t^{(1)}(2'3',23) d^{-1}(1) \delta(1',1) \\
t_3(1'2'3',123) &=& t^{(1)}(1'2',12) d^{-1}(3) \delta(3',3) . \eqn{t_3}
\eea

\bigskip
\bigskip
\centerline{\em 2. The $2\rightarrow 3$ process}
\bigskip

Although \eqs{T_i} and (\ref{T_2_sum}) determine the 1- and 2-particle
irreducible amplitude $T^{(2)}$, we still need to determine the amplitude
$M^{(2)}$ before the full $3\rightarrow 3$ amplitude $T$ can be specified by
\eq{T_a}.

Now to obtain the full connected $2\rightarrow 3$ amplitude $M_c$, we write it
in terms of one-particle irreducible and reducible parts:
\be
M_c = M_c^{(1)} + \Gamma^{(1)} g \bar{f} .
\ee
Similarly we can write
\be
M_c^{(1)} = M_c^{(2)} + \frac{1}{2} M^{(2)} D_0 t^{(1)} .
\ee
Using \eqs{Gamma_1} and (\ref{t}), we then obtain
\bea
M_c &=& M_c^{(2)} + M^{(2)} \frac{1}{2} D_0 t \nn
&=& M_d^{(2)} \frac{1}{2} D_0 t + M_c^{(2)} (1 + \frac{1}{2} D_0 t ) ,
\eea
where the disconnected part of $M^{(2)}$ is clearly given by
\be
M_d^{(2)}= \sum_{L_c} (F_1 + F_2 ) .    \eqn{M_d}
\ee
It is seen, therefore, that all we need now is the connected
amplitude $M_c^{(2)}$, and then both the coupled $3\rightarrow 3$ and
$2\rightarrow 3$ processes will be fully determined.

In the case of amplitude $M_c^{(2)}$, the very first possible three-particle
cut is not unique since the amplitude contains two-particles in the initial
state. However, here this non-uniqueness is simpler to handle than in the case
of
amplitudes with two-body intermediate states, since, apart from the two initial
particles, $M_c^{(2)}$ has no further two-particle states. In particular, we
can
avoid the overcounting problems associated with the non-uniqueness of the first
three-particle cut, by simply "pulling out" a vertex $f$ on one of
the two initial particles. Furthermore, the clearest way to see the structure
of  $M^{(2)}_c$ is  to temporarily expose the {\em bare} vertex function $f^0$
that is buried inside this chosen vertex $f$ of the skeleton diagram;
namely, we write \be
M^{(2)}_c = \frac{1}{2} [T^{(2)} G_0
F^0_1]_c   \eqn{M_c_TF}
\ee
where $F^0_1$ is defined as in \eq{F_i} but with
$f^0$ replacing $f$. \eq{M_c_TF} is illustrated in \fig{TF}.
\begin{figure}[b]
\vspace{2cm}
\caption{
\fign{TF}
Graphical representation of $M_c^{(2)}
=\frac{1}{2} [T^{(2)}G_0F_1^{0}]_c$.}
\thicklines
\begin{picture}(10,50)(-30,-10)

\put (100,0){\begin{picture}(100,100)
\put (5,105){$_{2'}$}
\put(5,75){$_{1'}$}
\put(5,90){$_{3'}$}
\put(110,105){$_{2}$}
\put(110,75){$_{1}$}

\multiput (20,75.)(0,30){2}{\line(1,0){80}}
\put (55,90){\oval(15,30)}
\put (62.5,90){\line(1,-1){15}}
\put (20,90){\line(1,0){27.5}}
\put (77,81){$f_0$}
\end{picture}}

\end{picture}
\end{figure}
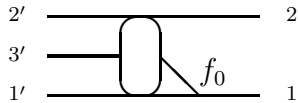
At this stage it is more straightforward to work in terms of Green's functions,
so we use \eq{G_12_T} to write the alternative form of \eq{M_c_TF},
\be
M^{(2)}_c = \frac{1}{2} [G^{-1}_0 G^{(2)} F^0_1]_c  . \eqn{M_c}
\ee
We also introduce the Green's functions $G_i$ defined by the equations
\bea
G_1 &=& D^{(1)}(1'3',13) g(2',2) \\
G_2 &=& D^{(1)}(2'3',23) g(1',1) \\
G_3 &=& D^{(1)}(1'2',12) g(3',3) .
\eea
Note that the $G_i$ are symmetric under the interchange of
labels corresponding to the interacting pair. It is easy to see that
\bea
G_i &=& G_0^{P_i} + \frac{1}{2}G_0 V_i G_i            \eqn{G_i}  \\
    &=& G_0^{P_i} + G_0 t_i G_0
\eea
where the symbol $P_i$ indicates a sum over permutations of (left or right)
momenta of the corresponding interacting pair:
\bea
G_0^{P_1} &\equiv& (1+P_{13}) G_0 = D_0^P(1'3',13) g(2',2) \\
G_0^{P_2} &\equiv& (1+P_{23}) G_0 = D_0^P(2'3',23) g(1',1) \\
G_0^{P_3} &\equiv& (1+P_{12}) G_0 = D_0^P(1'2',12) g(3',3) .
\eea
It is then evident that
\be
G_0 F_1 = \frac{1}{2} G_1 F^0_1 ,       \eqn{G_0 F_1}
\ee
and using \eq{G_i} to write $(1-\frac{1}{2}G_0V_1)G_1=G_0^{P_1}$, we obtain
\be
(1-\frac{1}{2}G_0V_1) G_0 F_1 = \frac{1}{2} G_0^{P_1} F_1^0 = G_0 F_1^0
\ee
from which follows that
\be
F_1^0 = (1-\frac{1}{2}V_1G_0)F_1 .
\ee
Using this result in \eq{M_c},
\be
M^{(2)}_c = \frac{1}{2} [G^{-1}_0 G^{(2)} (1-\frac{1}{2}V_1G_0) F_1]_c .
                           \eqn{M_c_1}
\ee
Writing the alternative form of \eq{G_12},
\be
G^{(2)} = G_0^P + G^{(2)} V G_0 ,
\ee
the previous expression becomes
\be
M^{(2)}_c = \frac{1}{2} G^{-1}_0 G^{(2)} (V_2/2+V_3/2+K_c/6) G_0 F_1 .
\eqn{M_c_2}
\ee
where the term $G^{-1}_0 G_0^PF_1$ has been dropped as it has no connected
part.
To obtain a scattering equation for $M^{(2)}_c$ we multiply \eq{M_c_2} on the
left by $1-VG_0$ and note that \eq{G_12} implies
\be
(G^{-1}_0 - V) G^{(2)} = G_0^{-1} G_0^P  = \sum_P \delta(1',1)
\delta(2',2) \delta(3',3) .    \eqn{GGP}
\ee
Thus
\be
 M^{(2)}_c = V G_0 M^{(2)}_c + \frac{1}{2} \sum_L (V_2/2+V_3/2+K_c/6)
G_0 F_1 . \eqn{M_c_LS}
\ee
With the derivation of this equation, we have, at least formally, solved the
double counting problem. \eq{M_c_LS} is an integral equation whose
solution provides $M_c^{(2)}$, and consequently, the full Green's functions
for both the $2\rightarrow 3$ and $3\rightarrow 3$ process; of course, this
equation has kernels $V_i$ which are non-compact, but this is an easily
overcome
formality. In a three-dimensional time-ordered approach, we would get, instead
of
\eq{M_c_LS}, the equation \be
 M^{(2)}_c = V G_0 M^{(2)}_c + \frac{1}{2} \sum_L (V_2/2+V_3/2+K_c/6)
G_0 F_1  + \frac{1}{2} \sum_L (V_1/2+V_3/2+K_c/6)
G_0 F_2, \eqn{M_c_3D}
\ee
which, in contrast to \eq{M_c_LS}, is symmetric with respect to the particle
labels 1 and 2. The seeming asymmetry of \eq{M_c_LS} implies that there must be
two-body forces, hidden within the three-body force $K_c$, that restore the
symmetry. Revealing these hidden two-body forces is therefore of great
conceptual importance, quite apart from questions of their importance in
practical calculations.

The amplitude $K_c$ is one-, two- and three-particle irreducible, and it is
connected. In this sense, it is a genuine three-body force and it would be
tempting to follow the common procedure of neglecting this term. This, however,
would be a major error: as illustrated in \fig{K_c F_1}, the combination $K_c
G_0 F_1$, entering \eq{M_c_LS}, in fact contains pair-like interactions when
viewed as having the very first vertex as $F_2$.
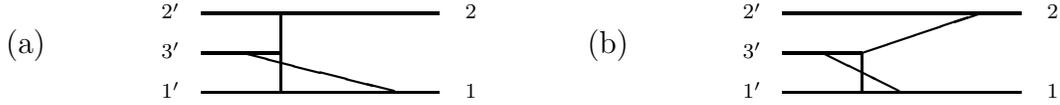
\begin{figure}[t]
\vspace{2cm}
\caption{\fign{K_c F_1} Two views of the same Feynman graph. (a) Viewed this
way, the
graph belongs to $K_c G_0 F_1$ where $K_c$ is a three-body force. (b) Viewed
this way, the graph belongs to $V_1 G_0 F_2$ where $V_1$ is a two-body
potential.}
\begin{picture}(0,10)(20,-10)
\put (0,0){ \begin{picture}(100,100)
\put (20,90){(a)}\end{picture}
\begin{picture}(100,100)(30,0)
\put (5,105){$_{2'}$}
\put (5,90){$_{3'}$}
\put (5,75){$_{1'}$}
\put (120,105){$_{2}$}
\put (120,75){$_{1}$}
\put (20,105.){\line(1,0){90.}}
\put (20,90){\line(1,0){30.}}
\put (20,75){\line(1,0){90.}}
\put (50,75){\line(0,1){30.}}
\put (35,90){\line(4,-1){58.}}
\end{picture}}

\put (220,0){ \begin{picture}(100,100)
\put (20,90){(b)}\end{picture}
\begin{picture}(100,100)(30,0)
\put (5,105){$_{2'}$}
\put (5,90){$_{3'}$}
\put (5,75){$_{1'}$}
\put (120,105){$_{2}$}
\put (120,75){$_{1}$}
\put (20,105.){\line(1,0){90.}}
\put (20,90){\line(1,0){30.}}
\put (20,75){\line(1,0){90.}}
\put (50,90){\line(3,1){45.}}
\put (50,75){\line(0,1){15.}}
\put (35,90){\line(2,-1){29.}}
\end{picture}}

\end{picture}
\end{figure}

Let us specify more precisely how this dual nature of the one diagram in
\fig{K_c F_1} manifests itself in our theory. To do this, we need to relate
our four-dimensional three-body problem to the corresponding formulation
in three-dimensional time-ordered perturbation theory. This can be done as
follows. Let ${\cal F}(p_1\ldots p_n, q_1\ldots q_m)$ be some Feynman
diagram contributing to the Green's function ${\cal G}$ of \eq{GG}, and
define $\tilde{{\cal F}}(E,\bfp_1\ldots\bfp_n,\bfq_1\ldots\bfq_m)$ by the
expression
\bea
\lefteqn{ \tilde{{\cal F}}(E,\bfp_1\ldots\bfp_n,\bfq_1\ldots\bfq_m) =
\int \delta(E-\sum_{i=1}^n p_i^0)\,dp_1^0\ldots dp_n^0}
\hspace{1cm}\nn
&&{\cal F}(p_1\ldots p_n,q_1\ldots q_m)
\delta(E-\sum_{i=1}^mq_i^0)\,dq_1^0\ldots dq_m^0    .             \eqn{convo}
\eea
Note that the integrations in \eq{convo} are in fact with respect to the
relative energies of the Feynman diagram. It can then be shown, that
$\tilde{\cal F}$ is the sum of all possible orderings of a time-ordered
perturbation graph topologically identical to ${\cal F}$, together with
additional terms due to the advanced part of the two-time Green's function
\cite{KB3}.  Thus if we take the
set of Feynman diagrams specified by $K_c G_0 F_1$, then
\bea
\lefteqn{ \widetilde{K_c G_0 F_1} = \int  \delta(E-p_1^0-p_2^0-p_3^0)
dp_1^0dp_2^0dp_3^0 }\nn
&& K_c G_0 F_1(p_1p_2p_3,q_1q_2) \delta(E-q_1^0-q_2^0) dq_1^0dq_2^0
\eea
includes the sum of all possible orderings of corresponding time-ordered
perturbation theory graphs; amongst these are diagrams whose graphical
representation is given by Figs. \ref{K_c F_1}(a) and (b). Now in time-ordered
perturbation theory, the diagram of  \fig{K_c F_1}(b), is unambiguously a
two-body rescattering in the presence of a spectator, and is therefore included
in any three-dimensional approach; indeed, neglecting this diagram
while keeping other two-body contributions would lead to violation of
three-body
unitarity. For these reasons it is of utmost importance that what we have
called
pair-like contributions in $K_c G_0 F_1$ be retained.

To examine more closely which pair-like diagrams of the form $V_i G_0 F_2$
($i=1,3$) also belong to the three-body force term $K_c G_0 F_1$, we expand
$V_i$
by order of the interaction: \be
V_i = V_i^{[2]} + V_i^{[4]} + \dots
\ee
where $V_i^{[n]}$ is the $n$-th order potential term. The example of
\fig{K_c F_1} shows already that the fourth-order term, $V_i^{[4]}$, gives
a diagram that is a member of  $K_c G_0 F_1$. Clearly, therefore,
all higher order terms of $V_i$ will likewise give diagrams belonging to
$K_c G_0 F_1$. This leaves only the lowest order term, $V_i^{[2]}$, as the
single exception as it does not belong to $K_c G_0 F_1$. The
corresponding terms  $V_1^{[2]} G_0 F_2$ and $V_3^{[2]} G_0 F_2$,
illustrated in \fig{fig_B_0}, are just appropriately symmetrized versions of
the
basic graph
\be
B(1'2'3',12) = \delta(1'2'3',12)
 d(1'') d(2'')  \, h(1'1'',1) h(-1''-2'',-2') h(3'2'',2) ,   \eqn{B}
\ee
shown as the bottom left diagram of \fig{fig_B_0}.
\begin{figure}[b]
\vspace{4.0cm}
\caption{
\fign{fig_B_0}
Terms of the form $V_iG_0F_2$ that do not belong to $K_{c}G_0F_1$.
The bottom left diagram is denoted by $B$ in the text. }
\begin{picture}(100,20)(-10,20)

\put (0,70){\begin{picture}(100,100)
\put (30,90){$V_1^{[2]}G_0F_2\,\,=\,$}\end{picture}
\begin{picture}(100,100)
\put (5,105){$_{2'}$}
\put (5,90){$_{3'}$}
\put (5,75){$_{1'}$}
\put (90,105){$_{2}$}
\put (90,75){$_{1}$}
\put (20,105.){\line(1,0){60.}}
\put (20,90){\line(1,0){30.}}
\put (20,75){\line(1,0){60.}}
\put (50,75){\line(0,1){30.}} \end{picture}}

\put (220,70){\begin{picture}(100,100)
\put (-16,87){$+$}
\put (5,105){$_{2'}$}
\put (5,90){$_{1'}$}
\put (5,75){$_{3'}$}
\put (90,105){$_{2}$}
\put (90,75){$_{1}$}
\put (20,105.){\line(1,0){60.}}
\put (20,90){\line(1,0){30.}}
\put (20,75){\line(1,0){60.}}
\put (50,75){\line(0,1){30.}} \end{picture}}

\put (0,10){\begin{picture}(100,100)
\put (30,90){$V_3^{[2]}G_0F_2\,\,=\,$}\end{picture}
\begin{picture}(100,100)
\put (5,105){$_{3'}$}
\put (5,90){$_{2'}$}
\put (5,75){$_{1'}$}
\put (90,105){$_{2}$}
\put (90,75){$_{1}$}
\put (20,105.){\line(1,0){60.}}
\put (55,95){$_{2''}$}
\put (55,80){$_{1''}$}
\put (20,90){\line(1,0){30.}}
\put (20,75){\line(1,0){60.}}
\put (50,75){\line(0,1){30.}} \end{picture}}

\put (220,10){\begin{picture}(100,100)
\put (-16,87){$+$}
\put (5,105){$_{3'}$}
\put (5,90){$_{1'}$}
\put (5,75){$_{2'}$}
\put (90,105){$_{2}$}
\put (90,75){$_{1}$}
\put (20,105.){\line(1,0){60.}}
\put (20,90){\line(1,0){30.}}
\put (20,75){\line(1,0){60.}}
\put (50,75){\line(0,1){30.}} \end{picture}}

\end{picture}
\end{figure}

Taking the above discussion into account, we can write
\be
K_c G_0  F_1 = \sum_{L_c} \left[ (V_1-V_1^{[2]})+(V_3-V_3^{[2]}) \right] G_0
F_2
+ K_{1c} G_0 F_1       \eqn{K_1c}
\ee
where $K_{1c}$ is made up of graphs which do not give rise to a two-body
interaction when multiplied from the right by $G_0 F_1$ with a subsequent
"pulling out" of vertex $F_2$.  Note that the combination $K_{1c} G_0 F_1$,
despite its notation, is completely symmetric; moreover,
diagrams belonging to $K_{1c} G_0 F_1$ do not give rise to pair-like
interactions when either $F_1$ or $F_2$ is pulled out to the right. This
suggests a possible classification of $2\rightarrow 3$ diagrams: (i) those that
give pair-like interactions when either of the two vertices is pulled out -
the
only possible case is shown in \fig{fig_B_0}, (ii) those that give pair-like
interactions when just one of the two vertices is pulled out; this case is
given
by \bea
\left[ (V_1-V_1^{[2]})+(V_3-V_3^{[2]}) \right] G_0 F_2 +
\left[ (V_2-V_2^{[2]})+(V_3-V_3^{[2]}) \right] G_0 F_1 , \nonumber
\eea
and (iii) those that do not give rise to pair-like interactions - $K_{1c} G_0
F_1$. Defining \be
B_0 = \frac{1}{2}(V_1^{[2]} + V_3^{[2]}) G_0 F_2,  \eqn{B_0}
\ee
and using \eq{K_1c}, we can write \eq{M_c_LS} as
\be
(1-VG_0) M^{(2)}_c
=  \frac{1}{2} \sum_{L} [\frac{1}{2}(V_2+V_3) G_0
F_1 + \frac{1}{2}(V_1+V_3) G_0 F_2 - B_0] + \frac{1}{2} K_{1c} G_0 F_1
\eqn{M_c_4}
\ee
where, because of the left permutation sum, $B_0$ can also be replaced by $2B$.
By symbolic form, this result differs from the corresponding result in
time-ordered perturbation theory, just in the subtraction of the term $B_0$.
This can be understood because, in our covariant case, the right most vertices
are not time-ordered relative to each other, so that  $B_0$ is contained in
both the first and second terms on the rhs of \eq{M_c_4}. Thus one subtraction
of $B_0$ is necessary to avoid overcounting. The potential $V$ on the lhs of
these equations still contains the three-body force  $K_c$ as given by \eq{V}.
Now, however, we can safely neglect this term as it is multiplied on the right
by a connected amplitude, and thus cannot give rise to pair-like interactions.
With the understanding that all genuine three-body forces are to be neglected,
\eq{M_c_4} provides a linear integral equation for the amplitude $M_c^{(2)}$
where all potentials are pair-like. We now need to perform a Faddeev
rearrangement in order to obtain a compact kernel. This we cannot do directly
to \eq{M_c_4} because of the sum over left permutations.

By analogy with \eq{T_tilde}, one way to proceed would be to define the
amplitude
$\tilde{M}$ satisfying the equation
\be
(1-VG_0) \tilde{M} = \frac{1}{2}[\frac{1}{2}(V_2+V_3) G_0 F_1
+ \frac{1}{2}(V_1+V_3) G_0 F_2 - 2B]                    \eqn{M_tilde}
\ee
where
\be
M^{(2)}_c = \sum_{L} \tilde{M},
\ee
so that \eq{M_tilde} is the same as \eq{M_c_4} but without the sum over
permutations. We may then express $\tilde{M}$ in terms of the set of equations
\bea
\tilde{M} &=& \sum_i \tilde{M}_i -B ,\\
\tilde{M}_i &=& V_iG_0 [ \frac{1}{2}\tilde{M}+\beta_i ] ,
\eea
where
\be
\beta_1=\frac{1}{4}F_2; \hspace{1cm}
\beta_2=\frac{1}{4}F_1; \hspace{1cm}
\beta_3=\frac{1}{4}(F_1+F_2).
\ee
A Faddeev rearrangement then gives the equations
\be
\tilde{M}_i = t_i G_0(\beta_i-\frac{1}{2}B) + \frac{1}{2}t_iG_0\sum_{j\ne i}
\tilde{M}_j.
\ee
This method corresponds to a reduction to the different particle  case;
however, as we now proceed to show, this reduction is not necessary as the sum
over permutations in \eq{M_c_4} can also be explicitly evaluated.

The evaluation of $\sum_{L} V_2 G_0 F_1$ and  $\sum_{L} V_3 G_0 F_1$ in
\eq{M_c_4}is most easily seen diagrammatically, as illustrated in \fig{V_2F_1}.
It is evident that
\thicklines
\begin{figure}[t]
\vspace{3.5cm}
\caption{
\fign{V_2F_1}
Graphical representation of the sums
(a) $\sum_{L} V_2 G_0 F_1 = 2V_2 G_0 F_1 + 2R_{12}V_1 G_0 F_2 + 2V_3 G_0 F_1$
and
(b) $\sum_{L} V_3 G_0 F_1 = 2V_3 G_0 F_1 + 2V_2 G_0 F_1 + 2R_{12}V_1 G_0 F_2$;
only the cyclic permutations of the left hand labels are shown.}
\begin{picture}(0,-155)(10,-80)  \put (0,90){(a)}
\put (50,0){ \begin{picture}(100,100)(30,0) \put
(5,105){$_{2'}$} \put (5,95){$_{3'}$} \put
(5,75){$_{1'}$} \put (62,110){$_{2''}$}
\put (62,91){$_{3''}$}
\put (62,68){$_{1''}$}

\put (110,105){$_{2}$}
\put (110,75){$_{1}$}
\put (20,105.){\line(1,0){80.}}
\put (20,95){\line(1,0){30.}}
\put (20,75){\line(1,0){80.}}
\put (50,95){\line(3,-2){30.}}
\put (50,100){\circle{10}}
\end{picture}}

\put (185,0){ \begin{picture}(0,0)(30,0)
\put (-15,88){$+$} \end{picture}
\begin{picture}(100,100)(30,0)
\put (5,105){$_{3'}$}
\put (5,95){$_{1'}$}
\put (5,75){$_{2'}$}
\put (62,110){$_{3''}$}
\put (62,91){$_{1''}$}
\put (62,68){$_{2''}$}

\put (110,105){$_{2}$}
\put (110,75){$_{1}$}
\put (20,105.){\line(1,0){80.}}
\put (20,95){\line(1,0){30.}}
\put (20,75){\line(1,0){80.}}
\put (50,95){\line(3,-2){30.}}
\put (50,100){\circle{10}}
\end{picture}}

\put (320,0){ \begin{picture}(0,0)(30,0)
\put (-15,88){$+$} \end{picture}
\begin{picture}(100,100)(30,0)
\put (5,105){$_{1'}$}
\put (5,95){$_{2'}$}
\put (5,75){$_{3'}$}
\put (62,110){$_{1''}$}
\put (62,91){$_{2''}$}
\put (62,68){$_{3''}$}

\put (110,105){$_{2}$}
\put (110,75){$_{1}$}
\put (20,105.){\line(1,0){80.}}
\put (20,95){\line(1,0){30.}}
\put (20,75){\line(1,0){80.}}
\put (50,95){\line(3,-2){30.}}
\put (50,100){\circle{10}}
\end{picture}}
\end{picture}

\begin{picture}(0,-35)(10,-30)
\put (0,90){(b)}
\put (50,0){ \begin{picture}(100,100)(30,0)
\put (5,105){$_{1'}$}
\put (5,95){$_{2'}$}
\put (5,75){$_{3'}$}
\put (62,110){$_{1''}$}
\put (62,91){$_{2''}$}
\put (62,68){$_{3''}$}

\put (110,105){$_{2}$}
\put (110,75){$_{1}$}
\put (20,105.){\line(1,0){80.}}
\put (20,95){\line(1,0){30.}}
\put (20,75){\line(1,0){80.}}
\put (50,95){\line(3,-2){30.}}
\put (50,100){\circle{10}}
\end{picture}}

\put (185,0){ \begin{picture}(0,0)(30,0)
\put (-15,88){$+$} \end{picture}
\begin{picture}(100,100)(30,0)

\put (5,105){$_{2'}$}
\put (5,95){$_{3'}$}
\put (5,75){$_{1'}$}
\put (62,110){$_{2''}$}
\put (62,91){$_{3''}$}
\put (62,68){$_{1''}$}

\put (110,105){$_{2}$}
\put (110,75){$_{1}$}
\put (20,105.){\line(1,0){80.}}
\put (20,95){\line(1,0){30.}}
\put (20,75){\line(1,0){80.}}
\put (50,95){\line(3,-2){30.}}
\put (50,100){\circle{10}}
\end{picture}}

\put (320,0){ \begin{picture}(0,0)(30,0)
\put (-15,88){$+$} \end{picture}
\begin{picture}(100,100)(30,0)
\put (5,105){$_{3'}$}
\put (5,95){$_{1'}$}
\put (5,75){$_{2'}$}
\put (62,110){$_{3''}$}
\put (62,91){$_{1''}$}
\put (62,68){$_{2''}$}

\put (110,105){$_{2}$}
\put (110,75){$_{1}$}
\put (20,105.){\line(1,0){80.}}
\put (20,95){\line(1,0){30.}}
\put (20,75){\line(1,0){80.}}
\put (50,95){\line(3,-2){30.}}
\put (50,100){\circle{10}}
\end{picture}}
\end{picture}

\end{figure}
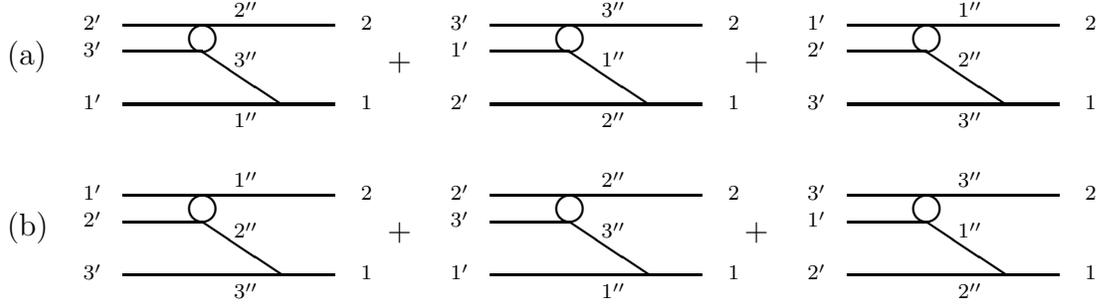


\be
\sum_{L} V_2 G_0 F_1 = \sum_{L} V_3 G_0 F_1.
\ee
In a similar way, we can consider the graphical representation of the terms
$\sum_{L} V_1 G_0 F_2$ and $\sum_{L} V_3 G_0 F_2$. In this way, we deduce that
\bea
\lefteqn{\sum_{L} [(V_2+V_3)G_0F_1 + (V_1+V_3)G_0F_2] } \hspace{4cm} \nn
&& = 4 V_1 G_0 F_2^R  + 4 V_2 G_0 F_1^R + 4 V_3 G_0 F_1^R   \eqn{V_nosum}
\eea
where, on the rhs, the potentials $V_i$ appear as in the different
particle case, without left-hand side permutation sums.

Substituting the last equation into \eq{M_c_4}, we obtain
\bea
\lefteqn{ (1-\frac{1}{2}\sum_i V_i G_0 -\frac{1}{6}K_cG_0) M_c^{(2)} }\nn
&&= V_1 G_0 F_2^R  + V_2 G_0 F_1^R + V_3 G_0 F_1^R -  B^L
+ \frac{1}{2}K_{1c}G_0F_1 .        \eqn{M_c_full}
\eea
For the description of the $2\rightarrow 3$ process, we can drop the terms
involving $K_c$ and $K_{1c}$ in this equation. It must be
emphasized, however, that once these three-body force terms are neglected, the
resulting amplitude $M_c^{(2)}$ should not be used to calculate the
$2\rightarrow 2 $ amplitude directly, i.e. through a multiplication on the left
by $\bar{F}_1 G_0$; such a procedure would certainly miss important
pair-like interactions coming from the term $\bar{F}_1 G_0 K_c$ and, as we
shall see shortly, also from the term  $\bar{F}_1 G_0 K_{1c}G_0F_1$. At this
stage we consider only the $2\rightarrow 3$ process so the neglect of $K_c$ and
$K_{1c}$ is justified. To lowest order, \eq{M_c_full} gives the expected result
$M_c^{(2)}=B^L$, thus providing a simple check of our result so far, and
illustrating the necessity of retaining the subtraction term $B^L$.

We can now perform the standard Faddeev rearrangement of \eq{M_c_full} by
defining
\bea
M^{(2)}_c &=& \sum_i M_i - B^L,                 \eqn{M_c_Faddeev}   \\
M_i &=& V_i G_0 [\frac{1}{2} M_c^{(2)} + \alpha_i]       \eqn{M_i}
\eea
where
\be
\alpha_1 = F_2^R ; \hspace{1cm} \alpha_2= F_1^R ; \hspace{1cm}
\alpha_3 = F_1^R \hspace{5mm} \mbox{or}\hspace{5mm} F_2^R.   \eqn{alpha}
\ee
With the help of \eq{t_i}, \eq{M_i} can be written in terms of
two-body t-matrices as
\be
M_i = t_i G_0 (\alpha_i - B^L) + \frac{1}{2} \sum_{j\ne i} t_i G_0 M_j .
\eqn{M_i_Faddeev}
\ee

\bigskip
\bigskip

\centerline{\em 3. The $2\rightarrow 2$ process}
\bigskip

In the above discussion, the $2\rightarrow 2$ amplitude $t$ was assumed to
be a known input. In the following discussion, $t$ will not be assumed to
be known, rather, we use the given field theoretic model to write an
integral equation for $t$.

The first steps, separating out the one- and two-particle cuts in $t$, have
already been performed in Eqs.\ (\ref{t}) and (\ref{t_(1)}). We therefore
begin by examining the structure of the $2\rightarrow 2$ potential $v$.
Since $v$ is connected and one-particle irreducible, both
of the final state particles in $v$  have separate left-most vertices. In
order to expose three-particle states, we shall "pull out" the left-most
vertex on the first particle. This is just the same procedure as was used
in \eq{M_c_TF} for exposing three-body states in $M^{(2)}$. We recall, that
in the case of $M^{(2)}$, we pulled out the bare vertex $f_0$. However, we
soon showed that this resulted in an expression, \eq{M_c_1}, where it is
the dressed vertex $f$ that has effectively been pulled out. This result
has its origin in the general property of quantum field theory where a careful
summation of Feynman diagrams leads to an effective perturbation theory where
only skeleton diagrams need be considered.

To underscore this equivalence of working with bare and dressed vertices, we
shall this time pull out the left-most dressed vertex on the first particle,
thus we write
\be v = \frac{1}{2} \bar{F}_1 G_0 M',
\ee
which is illustrated in \fig{FM}. Here $M'$ consists of all the diagrams of
$M^{(2)}$ except those that lead to dressing of vertices and particle
propagators in the above expression for $v$ (these restrictions are also
sufficient to guarantee the connectedness of $v$). Examples of such diagrams
belonging to $M^{(2)}$ but excluded from $M'$ are given in \fig{Bad_M}.
\begin{figure}[b]
\vspace{2cm}
\caption{
\fign{FM}
Graphical representation of $v=\frac{1}{2}
\bar{F}_1G_0M'.$}
\thicklines
\begin{picture}(10,50)(-30,-10)

\put (100,0){\begin{picture}(100,100)
\put (5,105){$_{2'}$}
\put(5,75){$_{1'}$}
\put(110,105){$_{2}$}
\put(110,75){$_{1}$}

\multiput (20,75.)(0,30){2}{\line(1,0){80}}
\put (65,90){\oval(15,30)}
\put (57.5,90){\line(-1,-1){15}}
\put (77,87){$M'$}
\end{picture}}

\end{picture}
\end{figure}
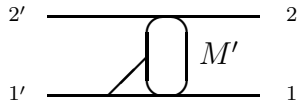

\thicklines
\begin{figure}[t]
\vspace{3cm}
\caption{
\fign{Bad_M}
Example of diagrams, belonging to $M^{(2)}$, that upon multiplication from the
left by $\bar{F}_1 G_0$, lead to dressing graphs. Diagram (a) leads to
propagator
dressing, (b) and (c) lead to vertex dressings.}
\begin{picture}(0,35)(35,-50)
\put (50,0){
\begin{picture}(100,100)(0,0)
\put (5,105){$_{2'}$} \put
(5,95){$_{3'}$} \put
(5,75){$_{1'}$} \put
(90,105){$_{2}$} \put
(90,75){$_{1}$} \put
(20,105.){\line(1,0){60.}} \put
(20,75){\line(1,0){60.}} \put
(50,75){\line(-3,2){30.}} \put
(43,50){(a)} \end{picture}}

\put (190,0){
\begin{picture}(100,100)
\put (5,105){$_{2'}$}
\put (5,90){$_{3'}$}
\put (5,75){$_{1'}$}
\put (90,105){$_{2}$}
\put (90,75){$_{1}$}
\put (20,105.){\line(1,0){60.}}
\put (20,90){\line(1,0){30.}}
\put (20,75){\line(1,0){60.}}
\put (50,75){\line(0,1){30.}}
\put (43,50){(b)}
\end{picture}}

\put (330,0){
\begin{picture}(100,100)
\put (5,105){$_{2'}$}
\put (5,90){$_{3'}$}
\put (5,75){$_{1'}$}
\put (90,105){$_{2}$}
\put (90,75){$_{1}$}
\put (20,105.){\line(1,0){60.}}
\put (20,90){\line(1,0){45.}}
\put (20,75){\line(1,0){60.}}
\put (50,90){\line(1,1){15.}}
\put (50,105){\line(1,-1){15.}}
\put (35,75){\line(0,1){15.}}
\put (65,75){\line(0,1){15.}}
\put (43,50){(c)}
\end{picture}}

\end{picture}
\end{figure}
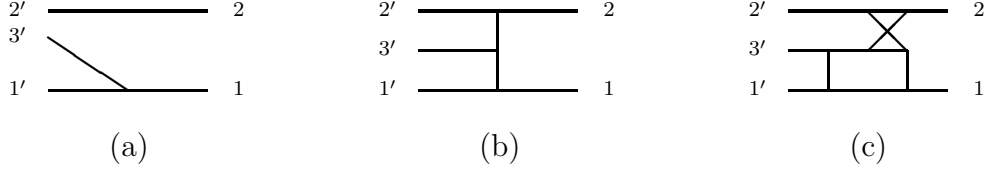


Writing $M'$ in terms of its connected and disconnected parts,
\be
v = \frac{1}{2} \bar{F}_1 G_0 M'_c + \frac{1}{2} \bar{F}_1 G_0 M'_d  \eqn{v_pr}
\ee
one can simply use Eqs.\ (\ref{M_c_full}) and (\ref{M_d}) for
$M^{(2)}_c$ and $M^{(2)}_d$ respectively, keeping only those terms that do not
give rise to propagator or vertex dressing; in this sense, we write \eq{v_pr}
as
\be
v = \frac{1}{2} \left\{ \bar{F}_1 G_0 M^{(2)}_c \right\}_{ND}
  + \frac{1}{2} \left\{ \bar{F}_1 G_0 M^{(2)}_d \right\}_{ND} \eqn{v_ND}
\ee
where subscript $ND$ stands for "no dressing". The part of $v$ resulting from
$M^{(2)}_d$ is thus given by
\be
v^{OPE} =  \frac{1}{2} \left\{ \bar{F}_1 G_0 M^{(2)}_d
\right\}_{ND}  = \bar{F}_1 G_0 F_2^R
\ee
which is just the properly symmetrized one-particle exchange (OPE) potential.
The full potential $v$ is thus
\bea
\lefteqn{v = v^{OPE} + \left\{ \frac{1}{2} \bar{F}_1 G_0
( V_2 G_0 F_1^R + V_3 G_0 F_1^R - \sum_L B + \frac{1}{2} K_{1c}
G_0 F_1) \right. } \hspace{3cm} \nn
 &&\left. + \frac{1}{2} \bar{F}_1 G_0 (\frac{1}{2} V_2 + \frac{1}{2} V_3
+ \frac{1}{6} K_c) G_0 M^{(2)}_c \right\}_{ND}. \eqn{vND_1}
\eea
To display the pair-interactions resulting from $K_c$, we use the analogue of
\eq{K_1c},
\be
\bar{F}_1 G_0 K_c = \sum_{R_c} \bar{F}_2 G_0 \left[
(V_1-V_1^{[2]})+(V_3-V_3^{[2]}) \right]  + \bar{F}_1 G_0 K'_{1c}   .
\eqn{K_1c_bar}
\ee
Note that $K'_{1c} \ne K_{1c}$. Moreover, $\bar{F}_1 G_0 K'_{1c}G_0M^{(2)}_c$
does not have pair-like rescattering contributions (with only one particle as
spectator) as can be easily seen by using the lowest order contribution for
$M^{(2)}_c$. Similarly, we define the analogues of Eqs.\ (\ref{B}) and
(\ref{B_0}) for the $3\rightarrow 2$ process by
\be
\bar{B}(1'2',123) = \delta(1'2',123)
 d(1'') d(2'')  \, h(1'1'',1) h(-1''-2'',2) h(2'2'',3) .   \eqn{B_bar}
\ee
and
\be
\bar{B}_0 = \frac{1}{2}\bar{F}_2 G_0 (V_1^{[2]} + V_3^{[2]})  \eqn{B_0_bar}
\ee
respectively. Using Eqs.\ (\ref{K_1c_bar}) and (\ref{B_0_bar}) in \eq{vND_1},
we obtain
\bea
\lefteqn{v = v^{OPE} + \frac{1}{2} \left\{ \bar{F}_1 G_0
( V_2 G_0 F_1^R + V_3 G_0 F_1^R - \sum_L B + \frac{1}{2} K_{1c} G_0 F_1)
\right. } \hspace{0cm} \nn
&& \left. +  [\frac{1}{2}\bar{F}_1 G_0 ( V_2 + V_3 )
+ \frac{1}{2}\bar{F}_2 G_0 ( V_1 + V_3 ) - \bar{B}_0
+ \frac{1}{6} \bar{F}_1 G_0K'_{1c}] G_0 M^{(2)}_c \right\}_{ND} .
\hspace{1cm} \eqn{vND_2}
\eea
At this stage, $M_c^{(2)}$ in this expression is still defined by the exact
expression of \eq{M_c_full}.
However, in contrast to the case of \eq{v_ND}, the factors multiplying
$M_c^{(2)}$ in \eq{vND_2} ensure that the three-body force
contributions to $M_c^{(2)}$ do not give rise to hidden two-body contributions
in \eq{vND_2}. We may therefore safely neglect such three-body contributions
and thus have $M_c^{(2)}$ described by \eq{M_c_Faddeev} rather than
\eq{M_c_full}.

In \eq{vND_2} terms of the form $V_iG_0M^{(2)}_c$ may be eliminated using Eqs.\
(\ref{M_i}) and (\ref{alpha}), and the symmetry of $M_c^{(2)}$ may be used to
reduce  $\bar{B}_0$ to $2\bar{B}$. Thus, neglecting "safe" three-body forces,
we obtain
\bea
\lefteqn{v = v^{OPE} + \frac{1}{2} \left\{
-  \bar{F}_1 G_0\sum_L B + \frac{1}{2}  \bar{F}_1 G_0 K_{1c} G_0 F_1
+  \bar{F}_1 G_0 M_2 + \bar{F}_2 G_0 M_1 \right. } \nn
&&\left.
+ (\bar{F}_1 + \bar{F}_2) G_0 M_3   - 2\bar{B} G_0 M^{(2)}_c -  \bar{F}_2 G_0
V_1 G_0 F_2^R  - \bar{F}_2 G_0 V_3 G_0 F_1^R \right\}_{ND}. \hspace{1cm}
\eqn{vND_3} \eea
We still need to identify  pair-like interactions that might be present in the
term involving the three-body force $K_{1c}$; that is, we seek hidden
expressions
of the form $\bar{F}_i G_0 V_k G_0 F_j$, including cases where the initial and
final momenta are permuted. In this respect, we note that all such pair-like
interactions can be expressed as one of the four terms $\bar{F}_i G_0 V_3 G_0
F_j$ $(i,j=1,2)$; for example, \bea \bar{F}_1G_0
V_2 G_0 F_1 &=& \bar{F}_1 G_0 V_3 G_0 F_1 ,\nn \bar{F}_2G_0 V_1 G_0 F_2 &=&
\bar{F}_2G_0 V_3 G_0 F_2 ,\nn L_{12} \bar{F}_1 G_0 V_3G_0 F_1 &=& \bar{F}_2 G_0
V_3G_0 F_1 ,\nn &\mbox{etc.}&    \nonumber
\eea
(of course cases like $\bar{F}_2 G_0 V_2 G_0 F_1$ do not arise as our
vertices are dressed from the beginning).
We note, therefore, that the last two terms of \eq{vND_3} may be simplified:
\bea
\frac{1}{2} (\bar{F}_2 G_0 V_1 G_0 F_2^R + \bar{F}_2 G_0 V_3 G_0 F_1^R )
&=& \bar{F}_2 G_0 V_3 G_0 (F_2+F_1) \\
&=& \bar{F}_2 G_0 V_3 G_0 F_2^R         \eqn{V3_R}
\eea
the result being illustrated in \fig{V3}. Note that this term is not symmetric
under the interchange of its left labels. This implies that there must be a
compensating term in \eq{vND_3} that restores the symmetry. Indeed we will see
that this compensating term can be found in  $\bar{F}_1 G_0 K_{1c} G_0 F_1$.
\thicklines
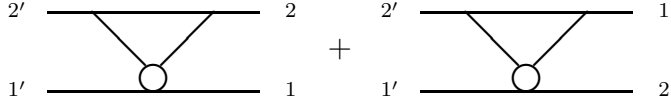
\begin{figure}[t]
\vspace{3cm}
\caption{
\fign{V3}
Graphical representation of $\bar{F}_2 G_0 V_3 G_0 F_2^R$.}

\begin{picture}(0,35)(35,0) \put (130,0){
\begin{picture}(100,100)(30,0)
\put (5,105){$_{2'}$}
\put (5,75){$_{1'}$}
\put (110,105){$_{2}$}
\put (110,75){$_{1}$}
\put (20,105.){\line(1,0){80.}}
\put (20,75){\line(1,0){80.}}
\put (37,105){\line(1,-1){20.}}
\put (83,105){\line(-1,-1){20.}}
\put (60,80){\circle{10}}
\end{picture}}
\put (275,0){
\put (-45,88){$+$}
\begin{picture}(100,100)(30,0)
\put (5,105){$_{2'}$}
\put (5,75){$_{1'}$}
\put (110,105){$_{1}$}
\put (110,75){$_{2}$}
\put (20,105.){\line(1,0){80.}}
\put (20,75){\line(1,0){80.}}
\put (37,105){\line(1,-1){20.}}
\put (83,105){\line(-1,-1){20.}}
\put (60,80){\circle{10}}
\end{picture}}
\end{picture}

\end{figure}

Central to the term $\bar{F}_1 G_0 K_{1c} G_0 F_1$ is the
three-body force $K_{1c}$. Although  $K_{1c}$ has been
defined [by \eq{K_1c}] so that no graph of $K_{1c}G_0F_1$ is
expressible as vertex $F_2$ followed by a two-body potential, the full term
$\bar{F}_1 G_0 K_{1c} G_0 F_1$, nevertheless, involves just such a
two-body potential. In \fig{FKF} we give an example of a graph, belonging to
$K_{1c}$, which, on being sandwiched between $\bar{F}_1 G_0$ and
$G_0 F_1$, is of the form $\bar{F}_2 G_0 V_3 G_0 F_2$. It is important that
such
pair-like interactions are not neglected in the theory.
\thicklines
\begin{figure}[b]
\vspace{3cm}
\caption{
\fign{FKF}
(a) Example of a graph belonging to the three-body force $K_{1c}$. (b) The
graph
of (a) sandwiched between operators $\bar{F}_1G_0$ and $G_0F_1$; the resulting
graph belongs to $\bar{F}_1G_0K_{1c}G_0F_1$, but is nevertheless of the
two-body
type $\bar{F}_2G_0V_3G_0F_2$.}
\begin{picture}(400,30)(20,-25)

\put (90,0){
\put (-70,88){(a)}
\begin{picture}(100,100)(30,0)
\put (5,105){$_{2'}$}
\put (5,90){$_{3'}$}
\put (5,75){$_{1'}$}
\put (110,105){$_{2}$}
\put (110,90){$_{3}$}
\put (110,75){$_{1}$}
\put (20,105.){\line(1,0){80.}}
\put (20,75){\line(1,0){80.}}
\put (70,90){\line(-3,2){22.}}
\put (50,90){\line(3,2){22.}}
\put (60,75){\line(0,1){15.}}
\put (20,90){\line(1,0){80.}}
\end{picture}}

\put (300,0){
\put (-70,88){(b)}
\begin{picture}(100,100)(30,0)
\put (5,105){$_{2'}$}
\put (5,75){$_{1'}$}
\put (110,105){$_{2}$}
\put (110,75){$_{1}$}
\put (20,105.){\line(1,0){80.}}
\put (20,75){\line(1,0){80.}}
\put (70,90){\line(-3,2){22.}}
\put (50,90){\line(3,2){22.}}
\put (40,75){\line(0,1){15.}}
\put (60,75){\line(0,1){15.}}
\put (80,75){\line(0,1){15.}}
\put (40,90){\line(1,0){40.}}
\end{picture}}
\end{picture}

\end{figure}
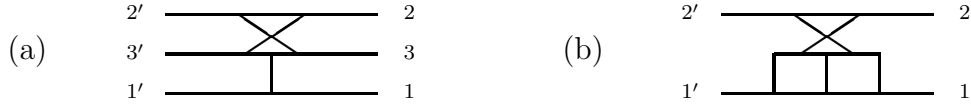

To identify all such pair-like interactions in $\bar{F}_1 G_0 K_{1c} G_0 F_1$,
it is sufficient to examine the four terms $\bar{F}_i G_0 V_3 G_0 F_j$
$(i,j=1,2)$. In this respect, we note that terms of the form $V_2G_0F_1$
or $V_3G_0F_1$ cannot belong to $K_{1c} G_0 F_1$ since $K_{1c}$ is a three-body
force.  Likewise, terms of the form $V_1G_0F_2$ or $V_3G_0F_2$ cannot belong to
$K_{1c} G_0 F_1$ just by the property of $K_{1c}$ given below \eq{K_1c}.
Consequently, out of the above mentioned four terms, two of them,
$\bar{F}_1 G_0 V_3G_0 F_1$ and $\bar{F}_1 G_0 V_3G_0 F_2$, cannot be of the
type
$\bar{F}_1 G_0 K_{1c} G_0 F_1$. This leaves only two candidates that are
consistent with the definition of $K_{1c}$, namely $\bar{F}_2 G_0  V_3 G_0 F_2$
and $\bar{F}_2 G_0 V_3 G_0 F_1$. However, even these have contributions which
do
not belong to $F_1 G_0 K_{1c} G_0 F_1$. In particular, we need to exclude from
$\bar{F}_2 G_0 V_3 G_0 F_2$ and $\bar{F}_2 G_0 V_3 G_0 F_1$ terms of the form
$\bar{F}_1 G_0 V_3G_0 F_1$ and $\bar{F}_1 G_0 V_3G_0 F_2$, as well as terms
corresponding to vertex dressing (that these are the only terms to be excluded
may be checked explicitly by using \eq{vND_3} to specify $V_3$). The only such
terms in our theory are given by the amplitudes $W$, $X$ and $Y$ defined by
 \fig{notFKF}, together with the same amplitudes but with the labels
on the two external right legs interchanged. The
diagram of \fig{notFKF}(a), defining amplitude $W$, corresponds to two-particle
exchange where the exchanged particles interact through the interaction $t'$,
where \be
t' = t^{(1)} - v^{OPE}.       \eqn{t'}
\ee
\thicklines
\begin{figure}[t]
\vspace{3cm}
\caption{
\fign{notFKF}
Graphs that need to be subtracted from the two-body rescattering term
$F_2G_0V_3G_0F_2$ as they do not belong to $F_1G_0K_{1c}G_0F_1$.
(a) Interacting two-particle exchange, $W$; in the s-channel, the
circle represents the amplitude $t'=t^{(1)}-v^{OPE}$. (b) Crossed two-particle
exchange , $X$. (c) Vertex dressing graph, $Y$.}
\begin{picture}(400,30)(10,0)

\put (30,60){
\begin{picture}(100,100)(30,0)
\put (5,105){$_{2'}$}
\put (5,75){$_{1'}$}
\put (110,105){$_{2}$}
\put (110,75){$_{1}$}
\put (20,105.){\line(1,0){80.}}
\put (20,75){\line(1,0){80.}}
\put (40,105){\line(4,-3){16.}}
\put (80,105){\line(-4,-3){16.}}
\put (40,75){\line(4,3){16.}}
\put (80,75){\line(-4,3){16.}}
\put (60,90){\circle{10}}
\put (52,48){(a)}
\end{picture}}

\put (180,60){
\begin{picture}(100,100)(30,0)
\put (5,105){$_{2'}$}
\put (5,75){$_{1'}$}
\put (110,105){$_{2}$}
\put (110,75){$_{1}$}
\put (20,105.){\line(1,0){80.}}
\put (20,75){\line(1,0){80.}}
\put (40,105){\line(4,-3){39.}}
\put (40,75){\line(4,3){39.}}
\put (52,48){(b)}
\end{picture}}

\put (330,60){
\begin{picture}(100,100)(30,0)
\put (5,105){$_{2'}$}
\put (5,75){$_{1'}$}
\put (110,105){$_{2}$}
\put (110,75){$_{1}$}
\put (20,105.){\line(1,0){80.}}
\put (20,75){\line(1,0){80.}}
\put (40,105){\line(4,-3){20.}}
\put (80,105){\line(-4,-3){20.}}
\put (60,75){\line(0,1){15.}}
\put (52,48){(c)}
\end{picture}}
\end{picture}

\end{figure}
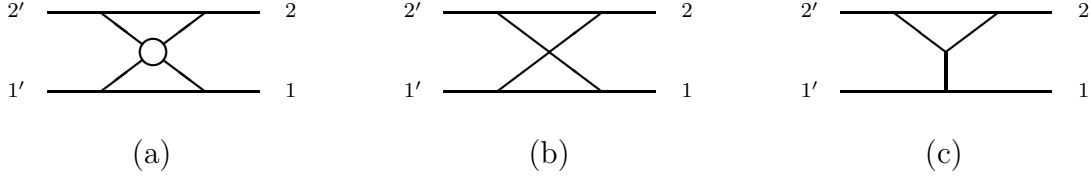
\noindent
Here the subtraction of $v^{OPE}$ is needed to eliminate contributions to
vertex
dressing. The diagram of \fig{notFKF}(b), defining amplitude $X$, corresponds
to
a "crossed two-particle exchange" potential; here the exchanged particles do
not
interact. The diagram of \fig{notFKF}(c), defining amplitude $Y$, corresponds
to
vertex dressing and must be excluded from our final expressions. In summary, we
can write that
\be
\frac{1}{4} \bar{F}_1 G_0 K_{1c} G_0 F_1 = \bar{F}_2 G_0 V_3 G_0 F^R_2
- W^R-2X-Y^R + \frac{1}{4} \bar{F}_1 G_0 K_{2c} G_0 F_1 .  \eqn{K2c}
\ee
where  $K_{2c}$ is a three-body force which when sandwiched between  $\bar{F}_1
G_0$ and $ G_0 F_1$, does not result in a pair-like interaction. Note that the
amplitude $X$ is symmetric and is thus equal to its exchange term; this and
other symmetry aspects of this expression are discussed in detail in the
Appendix.

On substituting the results of Eqs.\ (\ref{V3_R}) and (\ref{K2c}) into
\eq{vND_3}, the unsymmetrical terms of \fig{V3} are eliminated, and we obtain
\bea
\lefteqn{v = v^{OPE} + \left\{ \frac{1}{2} [ \bar{F}_1 G_0 M_2 + \bar{F}_2 G_0
M_1 + (\bar{F}_1 + \bar{F}_2) G_0 M_3 ] \right. } \hspace{2cm} \nn
&& \left. - \frac{1}{2} \bar{F}_1 G_0\sum_L B - \bar{B}G_0
M^{(2)}_c - W^R-2X-Y^R \right\}_{ND} \hspace{2cm}    \eqn{v_ND1}
\eea
where we have neglected the term
involving the three-body force $K_{2c}$. In this expression, we can further
evaluate
\be
\frac{1}{2} \bar{F}_1 G_0\sum_L B = X + (1+L_{12}R_{12})Y . \eqn{FB}
\ee

As written, \eq{v_ND1} specifies that dressing terms inside the curly brackets
need to be removed. This we can now do explicitly by identifying all possible
dressing terms, and then including subtraction terms in order to remove their
contribution. Some vertex dressing contributions have already been revealed in
the $Y$ amplitudes of Eqs.(\ref{v_ND1}) and (\ref{FB}). The only other vertex
dressings arise from the first iteration of the terms $\bar{F}_iG_0M_j$
in \eq{v_ND1}. Thus, with the $M_j$ evaluated to first order using
\eqs{M_i_Faddeev},
\bea
\lefteqn{\frac{1}{2}[\bar{F}_1G_0M_2+\bar{F}_2G_0M_1+(\bar{F}_1+\bar{F}_2)
G_0M_3]_{\,\mbox{(lowest order)}} }  \hspace{3cm} \nn &=&
\frac{1}{2}[\bar{F}_1G_0t_2G_0F_1^R + \bar{F}_2G_0t_1G_0F_2^R + (\bar{F}_1 +
\bar{F}_2)G_0t_3G_0F_1^R]\nn
&=& (\bar{F}_1+\bar{F}_2) G_0 t_3 G_0 (F_1+F_2)  = \sum_{LR} \bar{F}_2 G_0
t_3 G_0 F_2.
\eea
The vertex dressing terms arise from the one-particle exchange
potential part of $t_3$, that is,
\be
\sum_{LR} \bar{F}_2 G_0\, v_3^{OPE} G_0 F_2 = \sum_{LR}Y,
\ee
and these must be subtracted from \eq{v_ND1}.
Putting this together, the potential for the $2\rightarrow 2$ process is
\bea
\lefteqn{v = v^{OPE} + \frac{1}{2} [ \bar{F}_1 G_0 M_2 + \bar{F}_2 G_0 M_1
+ (\bar{F}_1 + \bar{F}_2) G_0 M_3 ]  - \bar{B} G_0 M^{(2)}_c } \hspace{6cm} \nn
&& - W^R-3X -\sum_{LR}Y  . \hspace{3cm}       \eqn{v_final}
\eea

\bigskip
\bigskip
\centerline{\bf B. The {\boldmath $\pi N\!N$} system}
\bigskip

Since the equations we have derived for identical particles contain all the
complications of the most general case, it is now essentially a formality
to write down the corresponding equations for other systems involving
three-point vertices. Here we would like to apply our formalism to the case of
the \piNN\  system which, because of its fundamental nature, has been under
intensive investigation for many years \cite{Garcilazo}. It is thus assumed
that we are dealing with a field theory of nucleons and pions where the
interaction is given by a \piNN\ vertex.

The discussion of the previous section does not need to be modified in any
essential way. However, one does need to take into account that now one has
only two identical particles, and  moreover, that these two are Fermions. In
this
respect, we need to modify our definitions of the various symmetry operations
introduced in the previous section. Although the symbols $L$ and $R$ still have
the meaning of sums over permutations, now the sums extend only over the two
nucleon labels, and with the additional condition that the "exchanged" term
enters with a minus sign. Choosing the convention that label $3$ always refers
to the pion, \eqs{A_R} and (\ref{A_L}) are now modified to
\bea
A^R(1'2'3',123) &=& A(1'2'3',123) - A(1'2'3',213) \\
A^L(1'2'3',123) &=& A(1'2'3',123) - A(2'1'3',123)
\eea
with similar relations holding for quantities having only nucleons in initial
or
final states.

\bigskip
\bigskip
\centerline{\em 1. $\pi N\!N \rightarrow \pi N\!N$ }
\bigskip

The amplitude $T$ for the $\pi N\!N \rightarrow \pi N\!N$ process is given,
as for the identical particle case, by \eq{T_a}. The 2-particle irreducible
amplitude $T^{(2)}$ is again the sum of Faddeev components $T_k$ as in
\eq{T_2_sum}; however, the equations for these components are now given by
\bea
&&T_i = t_i^R + t_i G_0 \sum_{k\ne i} T_k , \\
&&T_3 = t_3   + \frac{1}{2} t_3 G_0 \sum_{k\ne 3} T_k,
\eqn{T_3_piNN}
\eea
with
\bea
&&t_i = V_i +  V_i G_0 t_i , \\
&&t_3 = V_3 + \frac{1}{2} V_3 G_0 t_3 ,         \eqn{t_3_piNN}
\eea
where, in the above, $i=1,2$, and $k=1,2,3$. Note that $t_i$ is the \piN\
amplitude where the pion interacts with the $i$th nucleon, the other
nucleon being a spectator. Similarly $t_3$ is the \NN\ amplitude in the
presence of a spectator pion.

\bigskip
\bigskip
\centerline{\em 2. $N\!N \rightarrow \pi N\!N$}
\bigskip

With three-body forces neglected, the integral equation for the
$N\!N \rightarrow \pi N\!N$ 2-particle irreducible amplitude $M_c^{(2)}$ is
given by a slightly modified form of \eq{M_c_full}, namely
\be
M_c^{(2)} - (V_1 + V_2 + \frac{1}{2} V_3) G_0 M_c^{(2)}
= V_1 G_0 F_2^R  + V_2 G_0 F_1^R + V_3 G_0 F_1^R -  B^{LR}
\eqn{M_c_piNN}
\ee
where no $1/2$ factors appear in front of $V_1$ and $V_2$ since they
correspond to potentials between distinguishable particles (\piN ). The
amplitude $B$ now takes on the form shown in \fig{B_piN}, and sums
over both left and right nucleon exchanges are now also needed.

\begin{figure}[t]
\vspace{2.0cm}
\caption{
\fign{B_piN}
The term $B$ in the case of the $\pi N\!N$ system. }
\begin{picture}(100,20)(-10,20)

\put (150,0){\begin{picture}(100,100)
\put (5,105){$_{3'}$}
\put (5,90){$_{2'}$}
\put (5,75){$_{1'}$}
\put (90,90){$_{2}$}
\put (90,75){$_{1}$}
\put (20,90){\line(1,0){60.}}
\put (20,75){\line(1,0){60.}}
\multiput(40,75)(0,3.75){5}{\circle*{2}}
\multiput(65,90)(-4,1.25){12}{\circle*{2}}

\end{picture}}

\end{picture}
\end{figure}
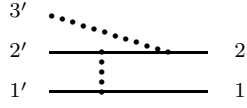

To effect the  Faddeev rearrangement of \eq{M_c_piNN} we define the
amplitude components by appropriately modified \eqs{M_i_Faddeev}
\bea
M_i &=& V_i G_0 [ M_c^{(2)} + F_j^R] ,  \hspace{1cm} (j\ne i)
                                                   \eqn{M_i_piNN} \\
M_3 &=& V_3 G_0 [\frac{1}{2} M_c^{(2)} + F_1^R] ,  \eqn{M_3_piNN} \\
M^{(2)}_c &=& \sum_k M_k - B^{LR}.
\eea
In the above equations, as in the two equations below, we use the convention
that $i,j=1,2$ while $k=1,2,3$. The rearranged equations are then given by
\bea
M_3 &=& t_3  G_0 (F_1^R - \frac{1}{2} B^{LR}) + \frac{1}{2} \sum_{k\ne 3}
t_3 G_0 M_k , \eqn{M_3} \\
M_i &=& t_i G_0 (F_j^R - B^{LR}) + \sum_{k\ne i}
t_i G_0 M_k \hspace{1cm} (j\ne i) .
\eea

\bigskip
\bigskip
\centerline{\em 3. $N\!N \rightarrow N\!N $}
\bigskip

For \NN\ scattering, there can be no contributions that are 1-particle
reducible
in the specified model where all interactions are generated by the \piNN\
vertex. The full \NN\ t-matrix, $t_{N\!N}$, is thus given by the Bethe-Salpeter
equation of \eq{t_(1)}, which in the present case takes the form
\be
t_{N\!N} = v_{N\!N} + \frac{1}{2} v_{N\!N} D_0 t_{N\!N} .    \eqn{t_NN}
\ee
Here $D_0$ is the dressed two-nucleon propagator, and the \NN\ potential,
$v_{N\!N}$, is specified in a similar way to \eq{v_final}, namely
\bea
\lefteqn{v_{N\!N} = v_{N\!N}^{OPE} + \bar{F}_1 G_0 M_2 + \bar{F}_2 G_0 M_1
 + (\bar{F}_1 + \bar{F}_2) G_0 M_3 }\hspace{1cm} \nn
&& - \bar{B}^{LR}  G_0 M^{(2)}_c - W_{\pi\pi}^R - W_{\pi N}^{LR}
- W_{N\!N}^R-X^R - Y^{LR} . \hspace{1cm}    \eqn{v_NN}
\eea
where $v_{N\!N}^{OPE}$ is now the \NN\ one-pion exchange potential and the
terms
$W_{\pi\pi}$, $W_{\pi N}$, $W_{N\!N}$, $X^R$ and $Y$ are illustrated in
\fig{WXY}. Each of the three different $W$ amplitudes involves the
exchange of two particles that interact with each other through
the appropriate version of the t-matrix $t'$ of \eq{t'}. Thus
the pions of amplitude $W_{\pi\pi}$ interact through the full $\pi$-$\pi$
t-matrix $t_{\pi\pi}$, the amplitude $W_{\pi N}$ involves $\pi$-$N$ scattering
through the so-called  background term, $t_{\pi N}^{(1)}$, and $W_{N\!N}$
involves \NN\ scattering through $t_{N\!N}-v_{N\!N}^{OPE}$. Note that
$W_{N\!N}$ does contain u-channel one-pion exchange, although t-channel
one-pion
exchange is forbidden.
\thicklines
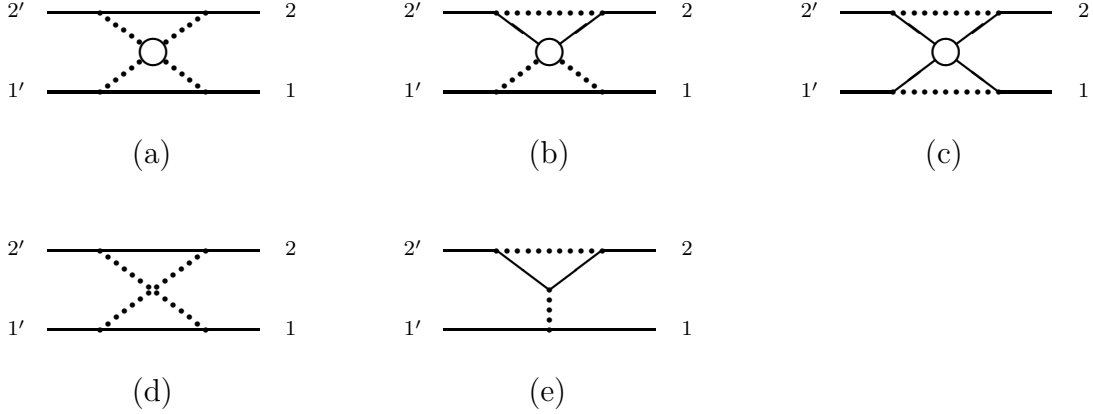
\begin{figure}[t]
\vspace{5cm}
\caption{
\fign{WXY}
Graphs that need to be subtracted from the $N\!N$ potential in the relativistic
formulation of the $\pi N\!N$ system.
(a) $W_{\pi\pi}$, (b)  $W_{\pi N}$,  (c)  $W_{N\!N}$, (d) $X$, and (e) $Y$.
The small circles represent the t-matrix amplitudes: (a) $t_{\pi\pi}$,
(b) $t_{\pi N}^{(1)}$, and (c) $t_{N\!N}-v_{N\!N}^{OPE}$, each being the
appropriate  version of \protect\eq{t'}. }
\begin{picture}(400,30)(10,0)

\put (30,140){
\begin{picture}(100,100)(30,0)
\put (5,105){$_{2'}$}
\put (5,75){$_{1'}$}
\put (110,105){$_{2}$}
\put (110,75){$_{1}$}
\put (20,105.){\line(1,0){80.}}
\put (20,75){\line(1,0){80.}}
\multiput(40,105)(3,-2.25){6}{\circle*{2}}
\multiput(80,105)(-3,-2.25){6}{\circle*{2}}
\multiput(40,75)(3,2.25){6}{\circle*{2}}
\multiput(80,75)(-3,2.25){6}{\circle*{2}}
\put (60,90){\circle{10}}
\put (52,48){(a)}
\end{picture}}

\put (180,140){
\begin{picture}(100,100)(30,0)
\put (5,105){$_{2'}$}
\put (5,75){$_{1'}$}
\put (110,105){$_{2}$}
\put (110,75){$_{1}$}
\put (20,105.){\line(1,0){20.}}
\put (80,105.){\line(1,0){20.}}
\put (20,75){\line(1,0){80.}}
\multiput(40,105)(4,0){11}{\circle*{2}}
\multiput(80,75)(-3,2.25){6}{\circle*{2}}
\multiput(40,75)(3,2.25){6}{\circle*{2}}
\put (40,105){\line(4,-3){16.}}
\put (80,105){\line(-4,-3){16.}}
\put (60,90){\circle{10}}
\put (52,48){(b)}
\end{picture}}

\put (330,140){
\begin{picture}(100,100)(30,0)
\put (5,105){$_{2'}$}
\put (5,75){$_{1'}$}
\put (110,105){$_{2}$}
\put (110,75){$_{1}$}
\put (20,105.){\line(1,0){20.}}
\put (80,105.){\line(1,0){20.}}
\put (20,75.){\line(1,0){20.}}
\put (80,75.){\line(1,0){20.}}
\multiput(40,105)(4,0){11}{\circle*{2}}
\multiput(40,75)(4,0){11}{\circle*{2}}
\put (40,105){\line(4,-3){16.}}
\put (80,105){\line(-4,-3){16.}}
\put (40,75){\line(4,3){16.}}
\put (80,75){\line(-4,3){16.}}
\put (60,90){\circle{10}}
\put (52,48){(c)}
\end{picture}}

\put (30,50){
\begin{picture}(100,100)(30,0)
\put (5,105){$_{2'}$}
\put (5,75){$_{1'}$}
\put (110,105){$_{2}$}
\put (110,75){$_{1}$}
\put (20,105.){\line(1,0){80.}}
\put (20,75){\line(1,0){80.}}
\multiput(40,105)(3.077,-2.308){14}{\circle*{2}}
\multiput(80,105)(-3.077,-2.308){14}{\circle*{2}}
\put (52,48){(d)}
\end{picture}}

\put (180,50){
\begin{picture}(100,100)(30,0)
\put (5,105){$_{2'}$}
\put (5,75){$_{1'}$}
\put (110,105){$_{2}$}
\put (110,75){$_{1}$}
\put (20,105.){\line(1,0){20.}}
\put (80,105.){\line(1,0){20.}}
\multiput(40,105)(4,0){11}{\circle*{2}}
\multiput(60,75)(0,3.75){5}{\circle*{2}}
\put (40,105){\line(4,-3){20.}}
\put (80,105){\line(-4,-3){20.}}
\put (20,75){\line(1,0){80.}}

\put (52,48){(e)}
\end{picture}}
\end{picture}

\end{figure}


It is important to note that the amplitude $t_{N\!N}$ enters our equations
both as an input and as an output. In this respect, we note that the
t-matrix $t_3$, defined formally by \eq{t_3}, in our case is just the
\NN\ t-matrix $t_{N\!N}$ in the presence of a spectator pion:
\be
t_3(1'2'3',123) =  t_{N\!N}(1'2',12) d_\pi^{-1}(3) \delta(3',3) .
\ee
Similarly, \eq{v_3} becomes in our case
\be
V_3(1'2'3',123) =  v_{N\!N}(1'2',12) d_\pi^{-1}(3) \delta(3',3) ,
\ee
and \eq{t_3_piNN} can thus be identified with \eq{t_NN}.
Since $t_3$ enters as an input to \eqs{T_3_piNN} and (\ref{M_3}), so does
$t_{N\!N}$. Similarly, $t_{N\!N}$ is needed as an input to calculate
$W_{N\!N}$.
At the same time, $t_{N\!N}$ is the output amplitude of the Bethe-Salpeter
equation \eq{t_NN}. Ideally, the input and output $t_{N\!N}$ need to be
self-consistent. This might be achievable through an iterative process where
the
output of one iteration becomes the input of the next iteration. Alternatively,
a great numerical simplification is afforded by sacrificing the
self-consistency
and using an externally constructed $t_{N\!N}$ for the input. This may be
reasonable if the observables of interest are not very sensitive to the
details of the \NN\ channel.

With the possible exception of $t_{N\!N}$, the two-body scattering t-matrices
entering the calculation of the $W$-amplitudes form part of the externally
constructed input to our equations.  It seems remarkable that the
$\pi$-$\pi$ t-matrix is a necessary input to our equations. No such input is
needed for the corresponding three-dimensional theory of the \piNN\ system, nor
has it appeared in previous formulations of four-dimensional \piNN\ equations.
In the present work, the appearance of this input arises
through the careful retention of all \piN\ interactions where one nucleon is a
spectator.

\bigskip
\bigskip
\centerline{\bf SUMMARY}
\bigskip

We have derived four-dimensional relativistic three-body equations based on
$\phi^3$ field theory. These equations couple the amplitudes for
$2\rightarrow 2$, $2\rightarrow 3$, and $3\rightarrow 3$ processes, and can
be considered as the natural extension of the Dyson-Schwinger and
Bethe-Salpeter equations to the three-particle sector. With the assumption
that three-body forces can be either neglected or given explicitly, these
three-body equations form a closed set of equations for the above
amplitudes, and provide the exact non-perturbative solution of the field
theoretical problem. In particular, these equations provide an answer to
the long standing question of how to determining the kernel of the
$2\rightarrow 2$ Bethe-Salpeter equation in a non-perturbative way.  If
three-body forces are not given, then in the spirit of our approach, one
could expose four-particle states in the three-body forces, and thus
express them in terms of four-particle equations where now four-body forces
are either neglected or explicitly given. Although one can, in this way,
envisage a whole hierarchy of equations, involving ever higher
multi-particle forces, there is a large amount of evidence that three-body
forces are small for most processes; thus, the three-body equations
developed in this paper, should form a solid theoretical starting point for
the investigation of relativistic few-body processes.

Two features distinguish our approach from previous attempts to formulate
the relativistic four-dimensional equations. Firstly, we have overcome the
overcounting problem, present in all previous works that use the so-called
"last-cut lemma"; to do this, we used a procedure of explicitly pulling out
one initial vertex in the $2\rightarrow 3$ amplitude.  Secondly, we have
retained all possible two-body interactions taking place in the presence of
a third spectator particle. In previous works, it was apparently not
realised that three-body forces in the $3\rightarrow 3$ amplitude can give
rise to such two-body forces in the $2\rightarrow 3$ amplitude. By
retaining appropriate three-body forces, we avoid the serious problem of
undercounting pair-like interactions which would arise if all three-body
forces in the $3\rightarrow 3$ amplitude were to be neglected.

Another interesting aspect of our four-dimensional equations is that they
differ at the "operator level" from the corresponding three-dimensional
three-body equations of time-ordered perturbation theory.  This is in contrast
to the case for the two-particle system where the four-dimensional
Bethe-Salpeter and three-dimensional Lippmann-Schwinger equations differ
essentially in the dimension of integrals involved, but otherwise have the same
operator form. For the three-body problem, this difference is due to the
presence of a number of subtractions terms in our four-dimensional equations
which are not present in the three-dimensional equations. These extra terms are
necessary to avoid overcounting and arise as a direct consequence of including
all the above-mentioned possible two-body interactions. We note that all
the previous works that suffer from overcounting [1-5] have four-dimensional
equations that are of the same operator form as the corresponding
three-dimensional equations.

It should be emphasised that our four-dimensional three-body equations are
not only of theoretical interest. With the ever increasing power of current
computers, these equations should also provide a timely practical tool for
calculating few-body hadronic processes in nuclear and high-energy physics.

\bigskip\bigskip
\centerline{\bf ACKNOWLEDGMENTS}
\bigskip

One of us (A.K.) would like to thank I.\ R.\ Afnan and A.\ W.\ Thomas for
the kind hospitality extended to him during his visit to Flinders and
Adelaide Universities.

\bigskip
\bigskip
\centerline{\bf APPENDIX}
\bigskip
\setcounter{equation}{0}
\renewcommand{\theequation}{A\arabic{equation}}

Here we would like to explain the origin of the counting factors appearing in
expressions of Section II dealing with identical particles.

The most common such factor is the inverse factorial $1/n!$ associated with
$n$-particle intermediate states connecting two amplitudes which are symmetric
with respect to label interchanges of these $n$ particles. Although this factor
can be derived by applying Wick's theorem to the Green's function of \eq{GG},
its appearance is not a feature of a four-dimensional approach, rather, it is
due to the basic symmetry property of identical particle amplitudes. Indeed,
just the same factor appears in three-dimensional approaches where, in fact,
the origin of this factor can be more easily seen. In a three-dimensional
approach, the Green's function corresponding to the four-dimensional one of
\eq{GG} is given by
\be
\tilde{\cal G} =\,\, <0|a(\bfp'_n)\ldots a(\bfp'_1)\frac{1}{E^+-H}
a^\dagger(\bfp_1)\ldots a^\dagger(\bfp_m)|0>       \eqn{GG_3d}
\ee
where $H$ is the Hamiltonian and the creation and annihilation operators
obey the commutation relations (for Bosons)
\be
[a(\bfp),a^\dagger(\bfp')] = \delta(\bfp-\bfp'); \hspace{1cm}
[a^\dagger(\bfp),a^\dagger(\bfp')] = [a(\bfp),a(\bfp')] = 0 .  \eqn{comm}
\ee
Just the same creation and annihilation operators appear in the expansion of
the
fields $\phi(x)$ in \eq{GG} (here there is no subscript on $\phi$ since all
particles are identical): \be
\phi(x) = \int \frac{d\bfp}{\sqrt{(2\pi)^3 2\omega_\bfp}}
[a(\bfp)e^{-ip\cdot x} + a^\dagger(\bfp)e^{ip\cdot x}] .
\ee
For three-dimensional Green's functions of \eq{GG_3d}, the states
\be
|\bfp_1\ldots\bfp_n> \,\, = a^\dagger(\bfp_1)\ldots a^\dagger(\bfp_n)|0>
\eqn{Fock}
\ee
form a basis of the full Fock space of identical particle states. As is well
known \cite{Negele}, the closure relation in this Fock space is given by
\be
|0><0| + \sum_{n=1}^\infty \int d\bfp_1\ldots d\bfp_n\, \frac{1}{n!}
|\bfp_1\ldots\bfp_n><\bfp_1\ldots\bfp_n| = 1 ,   \eqn{closure}
\ee
and here one can explicitly see the origin of the $1/n!$ factor for
$n$-particle
intermediate states.

We note that together with the condition $<0|0>\,\,=1$, \eqs{comm} or
\eq{closure} define the normalization of the states $|\bfp_1\ldots\bfp_n>$.
This normalization is particularly convenient for the case of disconnected
graphs as we now demonstrate. Let ${\cal G}_1(\alpha_i,\beta_i)$ and
${\cal G}_2(\alpha'_j,\beta'_j)$ denote two four-dimensional Green's functions
that make up the two disconnected parts of the disconnected
Green's function ${\cal G}_d(\alpha_i\alpha'_j,\beta_i\beta'_j)$. Here
we have used a short-hand notation for the momentum labels: e.g. $\alpha_i
\equiv \{\alpha_1,\alpha_2,\ldots\}$ denote the left-hand momenta of ${\cal
G}_1$
and $\alpha_i\alpha'_j \equiv
\{\alpha_1,\alpha_2,\ldots,\alpha'_1,\alpha'_2,\ldots\}$ denote the left-hand
momenta of ${\cal G}_d$. Note that ${\cal G}_1$, ${\cal G}_2$ and ${\cal G}_d$
are symmetric functions of their labels. Now it is straightforward to show that
\be
{\cal G}_d(\alpha_i\alpha'_j,\beta_i\beta'_j)
=\sum_{\stackrel{{\scriptstyle \alpha_i\leftrightarrow\alpha'_j}}
               {\beta_i \leftrightarrow \beta'_j}}
{\cal G}_1(\alpha_i,\beta_i) {\cal G}_2(\alpha'_j,\beta'_j)   \eqn{G_d_gen}
\ee
where the sum is over {\em distinct} terms resulting from all possible
exchanges of momentum labels, as indicated. The above mentioned
normalization ensures that this expression appears with no extra counting
factors. The result of \eq{G_d_gen} can be obviously generalized to
disconnected
Green's functions made up of any number of disconnected pieces, there again
being just a sum over exchange of labels with no extra counting factors.
In Section II, \eqs{sum_D_0}, (\ref{sum_G_0}), (\ref{K_d}) and (\ref{M_d}) can
be
considered as special cases of this result. It must be emphasized,
that although \eq{G_d_gen} is presented for four-dimensional Green's
functions, the structure of the result, in terms of a simple sum over label
exchanges, is due to the normalization and symmetry properties of
the underlying Fock states of \eq{Fock}. Thus the same equation holds for
three-dimensional Green's functions, although in that case, the simple product
of
${\cal G}_1$ and ${\cal G}_2$ in \eq{G_d_gen} needs to be replaced by a
convolution integral \cite{KB1}.

We now would like to discuss the various counting factors entering the
expression of \eq{K2c}. This will at the same time demonstrate how counting
factors can be determined for most other expressions.

The factor of $1/4$ entering the left side of \eq{K2c} is associated with the
vertex functions $\bar{F}_1$ and $F_1$ as they expose two identical particles
each.

Turning to the r.h.s. of \eq{K2c}, let us show that the amplitude $X$ of
\fig{notFKF}(b), constructed simply from four symmetric vertices and without
any
extra counting factors, is just the properly symmetrized non-interacting
two-particle exchange amplitude in our model. In \fig{XW}(a) we demonstrate how
such a symmetrized two-particle exchange amplitude is constructed. The factor
of
$1/16$ in \fig{XW}(a) results from factors of $1/2$ for each of the four
vertices
bracketing the symmetrized four-particle propagator. The factor of
$1/16$ could also have been obtained by first symmetrizing the products
$\bar{F}_1\bar{F}_2$ and $F_1F_2$ using \eq{G_d_gen}; in each case there would
be $6$ terms, each giving an equal contribution. One would then need to
multiply twice by $1/4!$, since a symmetrized product of two
vertices exposes $4$ particles in intermediate state. Thus $(6/4!)^2=1/16$,
as before. Now consider the sum over permutations $P$ of the
$4$-particle propagator in \fig{XW}(a): there will be 24 terms in this sum,
however, not all of them will lead to a two-particle exchange amplitude. In
fact it is easily seen that 8 of these permutations will result in disconnected
dressing graphs, leaving only $24-8=16$ that are proper two-particle exchanges.
Now all of these 16 contributions are equal to each other, so taking into
account the previous factor of 1/16, we obtain that the properly symmetrized
non-interacting two-particle exchange amplitude is just given by $X$, the
amplitude defined by the single diagram of \fig{notFKF}(b).
\begin{figure}[t]
\vspace{4cm}
\caption{
\fign{XW}
Illustration of how to determine the properly
symmetrized amplitudes for (a) non-interacting
two-particle exchange, and (b) interacting
two-particle exchange.  } \thicklines
\begin{picture}(100,0)(-30,0) \put
(100,70){\begin{picture}(100,100)
\put(-80,87){(a)} \put(-35,87){$X \,\, =\,\,
{\displaystyle \frac{1}{16}}$}

\put (90,105){$_{1}$}
\put(90,95){$_{2}$}
\put(90,85){$_{3}$}
\put(90,75){$_{4}$}

\multiput (20,75.)(0,30){2}{\line(1,0){30}}
\put (30,75){\line(2,1){20}}
\put (30,105){\line(2,-1){20}}
\multiput (150,75.)(0,30){2}{\line(1,0){30}}
\put (170,75){\line(-2,1){20}}
\put (170,105){\line(-2,-1){20}}
\put (70,80){\line(2,3){7}}
\put (70,101){\line(2,-3){7}}
\multiput (100,75.)(0,10){4}{\line(1,0){30}}
\put (65,90){\oval(10,50)[l]}
\put (135,90){\oval(10,50)[r]}
\put (70,67){$P$}
\put (70,80){\line(1,0){13}}
\put (70,101){\line(1,0){13}}
\end{picture}}

\put (100,0){\begin{picture}(100,100)
\put(-80,87){(b)}
\put(-35,87){$W \,\, =\,\,
{\displaystyle \frac{1}{16}}$}

\put(100,105){$_{1}$}
\put(100,95){$_{2}$}
\put(100,85){$_{3}$}
\put(100,75){$_{4}$}

\put(147,105){$_{1'}$}
\put(147,95){$_{2'}$}
\put(147,85){$_{3'}$}
\put(147,75){$_{4'}$}

\multiput (20,75.)(0,30){2}{\line(1,0){30}}
\put (30,75){\line(2,1){20}}
\put (30,105){\line(2,-1){20}}
\multiput (175,75.)(0,30){2}{\line(1,0){30}}
\put (195,75){\line(-2,1){20}}
\put (195,105){\line(-2,-1){20}}
\put (73,80){\line(1,0){13}}
\put (73,80){\line(2,3){7}}
\put (73,101){\line(2,-3){7}}
\put (73,101){\line(1,0){13}}
\multiput (110,75.)(0,10){4}{\line(1,0){30}}
\put (125,90){\circle{10}}
\put (65,90){\oval(10,50)[l]}
\put (160,90){\oval(10,50)[r]}
\put (70,67){$L'R'$}
\end{picture}}

\end{picture}
\end{figure}
In a similar way, let's find the symmetrization factor necessary in front of
the
interacting two-particle exchange amplitude $W$ defined in \fig{notFKF}(a). The
properly symmetrized interacting two-particle exchange is constructed as in
\fig{XW}(b). We again have the factor of $1/16$ due to the four vertices. This
time, however, two of the four particles in the center of the diagram interact.
In \fig{XW}(b), the symmetrization sum over $L'$ and $R'$ is, as in
\eq{G_d_gen}, over exchanges of left and right labels leading to {\em distinct}
contributions. Now out of the $4!\times 4!$ possible left and right
permutations, the switching of labels $2\leftrightarrow 3$, $2'\leftrightarrow
3'$ and $1'\leftrightarrow 4'$,  leads to a repetition of already included
terms, so they must be excluded. The sum over $L'$ and $R'$ thus extends over
only $4!4!/8$ terms. However, from all these terms, there will be cases where
the (interacting) legs $2$ and $3$ will both be connected to the same vertex,
thus leading to a vertex  dressing graph. Such cases must be excluded. Thus,
effectively, there are only 4 possibilities for legs $2$ and $3$: one leg
connected to either of the 2 legs of the top left  vertex, the other being
connected to either of the 2 legs of the bottom left  vertex. Similarly there
are 4 possibilities for the $2'$ and $3'$ legs of the middle
diagram to connect to the right-hand vertices. In total there can therefore be
only $4\times 4 = 16$ interacting two-particle exchanges, and taking into
account the initial $1/16$ factor, we see that the properly symmetrized
interacting two-particle exchange is given by the amplitude $W$ of
\fig{notFKF}(a) with no counting factors.

This kind of reasoning can be used for other types of diagrams involving
identical particles.

\setlength{\parsep}{-.1cm}

\end{document}